\documentclass[a4paper,11pt]{article}
\pdfoutput=1

\usepackage{jheppub} 

\usepackage[numbers,sort&compress]{natbib}
\usepackage{graphicx}  
\usepackage{float}
\usepackage{bm}        
\usepackage{amssymb}   
\usepackage{slashed}   
\usepackage{amsmath}   
\usepackage{verbatim}  
\usepackage{color} 
\usepackage{xcolor} 
\usepackage{subfig} 
\usepackage{epstopdf}

\definecolor{Blue}{rgb}{0, 0.1, 0.5}
\usepackage{hyperref}

\newcounter{JW}

\begin{document}

\title{Extract the energy scale of anomalous $\gamma\gamma \to W^+W^-$ scattering in the vector boson scattering process using artificial neural networks}
\author{Ji-Chong Yang}
\author{Jin-Hua Chen}
\author{Yu-Chen Guo}

\emailAdd{yangjichong@lnnu.edu.cn}
\emailAdd{cjh18241890611@163.com}
\emailAdd{ycguo@lnnu.edu.cn}
\affiliation{Department of Physics, Liaoning Normal University,\par No. 850 Huanghe Road, Dalian 116029, P.R. China}

\abstract{
As a model independent approach to search for the signals of new physics~(NP) beyond the Standard Model~(SM), the SM effective field theory~(SMEFT) draws a lot of attention recently.
The energy scale of a process is an important parameter in the study of an EFT such as the SMEFT.
However, for the processes at a hadron collider with neutrinos in the final states, the energy scales are difficult to reconstruct.
In this paper, we study the energy scale of anomalous $\gamma\gamma \to W^+W^-$ scattering in the vector boson scattering~(VBS) process $pp\to j j \ell^+\ell^-\nu\bar{\nu}$ at the large hadron collider~(LHC) using artificial neural networks~(ANNs).
We find that the ANN is a powerful tool to reconstruct the energy scale of $\gamma\gamma \to W^+W^-$ scattering.
The factors affecting the effects of ANNs are also studied.
In addition, we make an attempt to interpret the ANN and arrive at an approximate formula which has only five fitting parameters and works much better than the approximation derived from kinematic analysis.
With the help of ANN approach, the unitarity bound is applied as a cut on the energy scale of $\gamma\gamma \to W^+W^-$ scattering, which is found to has a significant suppressive effect on signal events.
The sensitivity of the process $pp\to j j \ell^+\ell^-\nu\bar{\nu}$ to anomalous $\gamma\gamma WW$ couplings and the expected constraints on the coefficients at current and possible future LHC are also studied.
}

\maketitle

\section{\label{level1}Introduction}

The Standard Model~(SM) has proven to be very successful and accurate.
Searching for new physics~(NP) beyond the SM is one of the main goals of current and future colliders.
Due to the lack of clear guidelines, a model independent approach to look for NP signals has gradually become popular, known as the SM effective field theory~(SMEFT)~\cite{weinberg,SMEFTReview1,SMEFTReview2,SMEFTReview3}.
It is assumed that energy scales of processes at current colliders are not large enough to directly produce the signals of NP particles.
At low energies, the NP sector is decoupled, one can integrate out NP particles, then NP effects become new interactions of known particles, which are in the form of higher dimensional operators.
Then, the SM can be extended as a low energy EFT of some unknown UV completion by adding those higher dimensional operators with small Wilson coefficients, result in a Lagrangian as
\begin{equation}
\begin{split}
&\mathcal{L}_{\rm SMEFT}=\mathcal{L}_{SM}+\sum _i\frac{C_{6i} }{\Lambda^2}\mathcal{O}_{6i}+\sum _j\frac{C_{8j}}{\Lambda^4}\mathcal{O}_{8j}+\ldots,
\end{split}
\label{eq.1.1}
\end{equation}
where $\mathcal{O}_{6i}$ and $\mathcal{O}_{8j}$ are dimension-6 and dimension-8 operators, $C_{6i}/\Lambda ^2$ and $C_{8j}/\Lambda ^4$ are corresponding Wilson coefficients, $\Lambda$ is the energy scale of NP.
In Eq.~(\ref{eq.1.1}), we have neglected the odd-dimensional operators which violate the lepton number conservation.

To investigate an EFT, the energy scale is an important parameter, because the Wilson coefficients are functions of energy scales.
It has been suggested that, in experiments the constraints on the Wilson coefficients of higher dimensional operators should be given as functions of energy scales~\cite{matchingidea1}.
Meanwhile, there are theoretical constraints such as the unitarity bounds~\cite{unitarityHistory1,unitarityHistory2,unitarityHistory3,partialwaveunitaritybound,jrr1} which are also functions of energy scales.
In conclusion, the reconstruction of the center-of-mass~(c.m.) energy is an important task in phenomenological studies of the SMEFT.

At a proton-proton~(pp) collider such as the Large Hadron Collider~(LHC), due to the parton distribution function~(PDF), the c.m. energy can only be reconstructed by using the information in the final states.
This poses difficulties for processes whose final states contain neutrinos.
For example, for the vector boson scattering~(VBS) process $pp\to jj \gamma W$, in order to study the unitarity bounds of anomalous quartic couplings~(aQGCs), one needs to reconstruct the c.m. energy of subprocess $\gamma(Z)W\to \gamma W$ subjected to a delicate kinematic analysis and approximation~\cite{wastudy}.
Another example is the process $pp\to WW$, where the study of validity of the SMEFT has also encountered great difficulties due to the neutrinos in the final state~\cite{atgcsuppresed,efttraingle3}.

There is a similar problem in the studies of processes containing vector bosons at the LHC.
The longitudinal polarized vector bosons are related to the symmetry broken and the Higgs mechanism, therefore draws a lot of attention~\cite{wpolarization,wpolarizationexp1,wpolarizationexp2,wpolarizationexp3,wpolarizationexp4,wpolarizationexp5}.
The polarization of a vector boson can be inferred by the momentum of the daughter charged lepton in the rest-frame of the vector boson, the so called helicity frame~\cite{wfraction}.
However, the momentum of the $W^{\pm}$ boson is difficult to reconstruct due the neutrino, as a result, it is difficult to boost the charged lepton to the rest frame of the $W$ boson, which is one of the reasons that the polarization of a $W$ boson is difficult to determinate.
In response to the problems in determining the polarizations, a novel approach has been introduced into high energy physics~(HEP).
It has been shown that, the artificial neural network~(ANN) can be very powerful in determining the polarizations of $W,Z$ bosons~\cite{wpolarizationANN1,wpolarizationANN2,zpolarizationANN} and $\tau$ lepton~\cite{taupolarizationANN}.
The ANN approach is one of the machine learning methods, which have been widely used in HEP, and are being developed rapidly in recent years~\cite{annhep1,annhep2,annhep3,mlreview,ml1,ml2,ml3,ml4,ml5,ml6}.

In this paper, we study the aQGCs induced by dimension-8 operators~\cite{aqgcold,aqgcnew} in the process $pp\to jj W^+W^-$ with leptonic decays of $W^{\pm}$ bosons.
The aQGCs can be contributed by a lot of NP models~\cite{bi1,bi2,composite1,composite2,extradim,2hdm1,2hdm2,zprime1,zprime2,alp1,alp2}, and has been studied intensively~\cite{aqgc1,aqgc2,vbscan,za,aaww,5gaugeinteraction}.
Dimension-6 operators cannot contribute to aQGCs while leaving anomalous triple gauge couplings~(aTGCs) along~\cite{vbscan}, therefore we concentrate on the dimension-8 operators.
A recent study shows that the existence of dimension-8 operators is necessary as long as the dimension-6 operators exist in the convex geometry point of view to the SMEFT space~\cite{convexgeometry}.
Besides, there are cases that the contributions from dimension-6 operators are absent~\cite{bi1,bi2,ntgc1,ntgc2,ntgc3,ntgc4,ntgc5}.
Moreover, aQGCs can lead to richer helicity combinations than dimension-6 aTGCs~\cite{ssww}.
Apart from that, aQGCs can be generated by tree diagrams while aTGCs are generated by loop diagrams~\cite{looportree}, therefore the possibility exists that the signals of dimension-8 aQGCs are more significant than the dimension-6 aTGCs.
Consequently, while the SMEFT has mainly been applied with dimension-6 operators, recently the study of dimension-8 operators has gradually received much attention~\cite{aqgcold,aqgcnew,ntgc1,d8}.
The most sensitive processes for aQGCs are the VBS processes~\cite{vbsreview}.
The VBS processes have been extensively studied by both the ATLAS and the CMS groups~\cite{sswwexp1,ssww,zaexp1,zaexp2,zaexp3,waexp1,zzexp1,zzexp2,wzexp1,wzexp2,wwexp2,coefficient1,waconstraint,eventclipping1,zzjjexp1,zajjexp2}, and will continue to draw attentions with future runs of the LHC.
The evidence of exclusive or quasi-exclusive $\gamma \gamma \to W^+W^-$ process has been found~\cite{wwexp1}.
The next-to-leading order QCD corrections to the process $pp\to W^+W^- jj$ have been computed~\cite{vbfcut}, and the $K$ factor is found to be close to one ($K\approx 0.98$).
As introduced, to study the dimension-8 operators, the two neutrinos in the final state will cause difficulties.
However, these difficulties just provide a good test for the ANN approach.
We use the ANN approach to study the process $pp\to jj \ell^+\ell^-\nu\bar{\nu}$ with the focus on the reconstruction of the energy scale of the $\gamma\gamma \to W^+W^-$ subprocess.
We discuss the justification of using ANN to study the energy of the subprocess, and show that the ANN can achieve better results than kinematic analysis.
An interpretation of the ANN is discussed, which indicates that the ANN can be approximated by a function of three variables and contains five fitting parameters.
The unitarity bounds and the signal significances of the aQGCs are also studied in this paper.

The remainder of the paper is organized as follows, in Sec.~\ref{level2} we briefly introduce the aQGCs; in Sec.~\ref{level3} the kinematic analysis is presented; the numerical results of the ANN approach is shown in Sec.~\ref{level4}; an interpretation of the ANN is presented in Sec.~\ref{level5}; in Sec.~\ref{level6}, we use the results of ANN to study the unitarity bounds and signal significances of aQGCs; Sec.~\ref{level7} is a summary.

\section{\label{level2}A brief introduction of aQGCs}

In this section, we briefly introduce the dimension-8 operators contributing to the aQGCs frequently used in experiments.
The Lagrangian relevant to the process $\gamma\gamma \to W^+W^-$ is $\mathcal{L}_{\rm aQGC}=\sum _{i} (f_{S_i}/\Lambda^4)O_{S,i}+\sum _{j} (f_{M_j}/\Lambda^4)O_{M,j}+\sum _{k} (f_{T_k}/\Lambda^4)O_{T,k}$ with~\cite{aqgcold,aqgcnew}
\begin{equation}
\begin{array}{ll}
O_{S,0}=\left[\left(D_{\mu}\Phi \right) ^{\dagger} D_{\nu}\Phi\right]\times \left[\left(D^{\mu}\Phi \right) ^{\dagger} D^{\nu}\Phi\right],
 &O_{T,0}={\rm Tr}\left[\widehat{W}_{\mu\nu}\widehat{W}^{\mu\nu}\right]\times {\rm Tr}\left[\widehat{W}_{\alpha\beta}\widehat{W}^{\alpha\beta}\right],\\
O_{S,1}=\left[\left(D_{\mu}\Phi \right) ^{\dagger} D_{\mu}\Phi\right]\times \left[\left(D^{\nu}\Phi \right) ^{\dagger} D^{\nu}\Phi\right],
 &O_{T,1}={\rm Tr}\left[\widehat{W}_{\alpha\nu}\widehat{W}^{\mu\beta}\right]\times {\rm Tr}\left[\widehat{W}_{\mu\beta}\widehat{W}^{\alpha\nu}\right],\\
O_{S,2}=\left[\left(D_{\mu}\Phi \right) ^{\dagger} D_{\nu}\Phi\right]\times \left[\left(D^{\nu}\Phi \right) ^{\dagger} D^{\mu}\Phi\right],
 &O_{T,2}={\rm Tr}\left[\widehat{W}_{\alpha\mu}\widehat{W}^{\mu\beta}\right]\times {\rm Tr}\left[\widehat{W}_{\beta\nu}\widehat{W}^{\nu\alpha}\right],\\
O_{M,0}={\rm Tr\left[\widehat{W}_{\mu\nu}\widehat{W}^{\mu\nu}\right]}\times \left[\left(D^{\beta}\Phi \right) ^{\dagger} D^{\beta}\Phi\right],
 &O_{T,5}={\rm Tr}\left[\widehat{W}_{\mu\nu}\widehat{W}^{\mu\nu}\right]\times B_{\alpha\beta}B^{\alpha\beta},\\
O_{M,1}={\rm Tr\left[\widehat{W}_{\mu\nu}\widehat{W}^{\nu\beta}\right]}\times \left[\left(D^{\beta}\Phi \right) ^{\dagger} D^{\mu}\Phi\right],
 &O_{T,6}={\rm Tr}\left[\widehat{W}_{\alpha\nu}\widehat{W}^{\mu\beta}\right]\times B_{\mu\beta}B^{\alpha\nu},\\
O_{M,2}=\left[B_{\mu\nu}B^{\mu\nu}\right]\times \left[\left(D^{\beta}\Phi \right) ^{\dagger} D^{\beta}\Phi\right],
 &O_{T,7}={\rm Tr}\left[\widehat{W}_{\alpha\mu}\widehat{W}^{\mu\beta}\right]\times B_{\beta\nu}B^{\nu\alpha},\\
O_{M,3}=\left[B_{\mu\nu}B^{\nu\beta}\right]\times \left[\left(D^{\beta}\Phi \right) ^{\dagger} D^{\mu}\Phi\right],
& O_{T,8}=B_{\mu\nu}B^{\mu\nu}\times B_{\alpha\beta}B^{\alpha\beta},\\
O_{M,4}=\left[\left(D_{\mu}\Phi \right)^{\dagger}\widehat{W}_{\beta\nu} D^{\mu}\Phi\right]\times B^{\beta\nu},
& O_{T,9}=B_{\alpha\mu}B^{\mu\beta}\times B_{\beta\nu}B^{\nu\alpha},\\
O_{M,5}=\left[\left(D_{\mu}\Phi \right)^{\dagger}\widehat{W}_{\beta\nu} D_{\nu}\Phi\right]\times B^{\beta\mu} + h.c., & \\
O_{M,7}=\left(D_{\mu}\Phi \right)^{\dagger}\widehat{W}_{\beta\nu}\widehat{W}_{\beta\mu} D_{\nu}\Phi, & \\
\end{array}
\label{eq.2.1}
\end{equation}
where $\Phi$ is the SM Higgs doublet, $\widehat{W}\equiv \vec{\sigma}\cdot \vec{W}/2$ with $\sigma$ being the Pauli matrices and $\vec{W}\equiv \{W^1, W^2, W^3\}$.
The $O_{M_{0,1,2,3,4,5,7}}$ and $O_{T_{0,1,2,5,6,7}}$ operators can contribute to five anomalous $\gamma\gamma WW$ couplings, which can be written as $\mathcal{L}_{\gamma\gamma WW}=\sum _{i=0}^4 \alpha_i V_{i}$ with~\cite{aaww}
\begin{equation}
\begin{array}{ll}
V_{0}=F_{\mu\nu}F^{\mu\nu}W^{+\alpha}W^-_{\alpha}, &V_{1}=F_{\mu\nu}F^{\mu\alpha}W^{+\nu}W^-_{\alpha},\\
V_{2}=F_{\mu\nu}F^{\mu\nu}W^+_{\alpha\beta}W^{-\alpha\beta}, &V_{3}=F_{\mu\nu}F^{\nu\alpha}W^+_{\alpha\beta}W^{-\beta\mu},\\
V_{4}=F_{\mu\nu}F^{\alpha\beta}W^{+\mu\nu}W^-_{\alpha\beta},& \\
\end{array}
\label{eq.2.2}
\end{equation}
where $W^{\pm}_{\mu\nu}\equiv \partial _{\mu}W^{\pm}_{\nu}-\partial _{\nu}W^{\pm}_{\mu}$.
The coefficients of the couplings can be related to the coefficients of the operators as
\begin{equation}
\begin{array}{ll}
\alpha_0=\frac{e^2v^2}{8\Lambda ^4}\left(f_{M_0}+\frac{c_W}{s_W}f_{M_4}+2\frac{c_W^2}{s_W^2}f_{M_2}\right),
&\alpha_1=\frac{e^2v^2}{8\Lambda ^4}\left(\frac{1}{2}f_{M_7}+2\frac{c_W}{s_W}f_{M_5}-f_{M_1}-2\frac{c_W^2}{s_W^2}f_{M_3}\right),\\
\alpha_2=\frac{1}{\Lambda ^4}\left(s_W^2f_{T_0}+c_W^2f_{T_5}\right),\;
&\alpha_3=\frac{1}{\Lambda ^4}\left(s_W^2f_{T_2}+c_W^2f_{T_7}\right),\;\\
\alpha_4=\frac{1}{\Lambda ^4}\left(s_W^2f_{T_1}+c_W^2f_{T_6}\right). &
\end{array}
\label{eq.2.3}
\end{equation}
Because each dimension-8 operator contributes to only one vertex, and the constraints on the dimension-8 operators are obtained by assuming one operator at a time in experiments, the constraints on $\alpha _i$ can be derived by the constraints on dimension-8 operators~\cite{aaww} which are listed in Table~\ref{tab.1}.
For simplicity, we concentrate on these five couplings in this paper.

\begin{table}[!htbp]
\centering
\begin{tabular}{cc|cc}
\hline
vertex & constraint & coefficient & constraint\\
\hline
$\alpha_0 ({\rm TeV^{-2}})$ & $[-0.013, 0.013]$ & $f_{M_2}/\Lambda ^4\;({\rm TeV^{-4}})$ & $[-2.8, 2.8]$ \\
$\alpha_1 ({\rm TeV^{-2}})$ & $[-0.021, 0.021]$ & $f_{M_5}/\Lambda ^4\;({\rm TeV^{-4}})$ & $[-8.3, 8.3]$ \\
$\alpha_2 ({\rm TeV^{-4}})$ & $[-0.38, 0.38]$ & $f_{T_5}/\Lambda ^4\;({\rm TeV^{-4}})$ & $[-0.5, 0.5]$ \\
$\alpha_3 ({\rm TeV^{-4}})$ & $[-0.69, 0.69]$ & $f_{T_7}/\Lambda ^4\;({\rm TeV^{-4}})$ & $[-0.9, 0.9]$ \\
$\alpha_4 ({\rm TeV^{-4}})$ & $[-0.31, 0.31]$ & $f_{T_6}/\Lambda ^4\;({\rm TeV^{-4}})$ & $[-0.4, 0.4]$ \\
\hline
\end{tabular}
\caption{\label{tab.1}The constraints on anomalous $\gamma\gamma W W$ couplings and the corresponding limits on the dimension-8 operators at 95\% CL~\cite{waconstraint}.
}
\end{table}

\begin{figure}[!htbp]
\centering{
\includegraphics[width=0.7\textwidth]{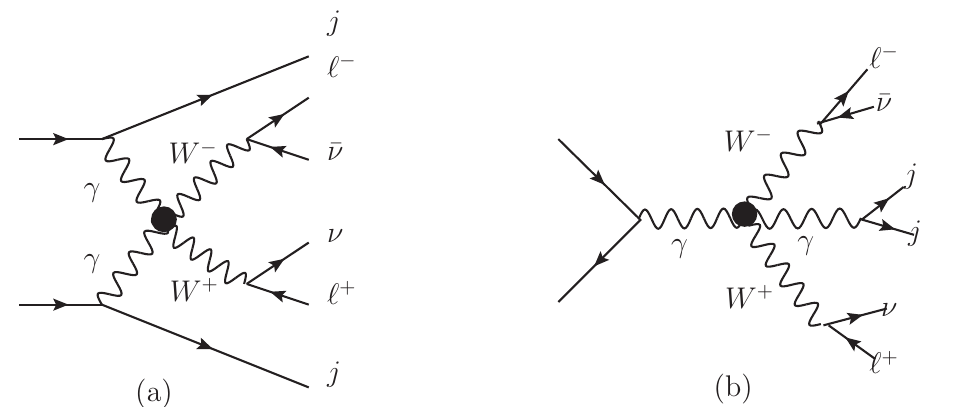}
\caption{\label{Fig:feynman1}
The Feynman diagrams of the contributions from anomalous $\gamma\gamma W W$ couplings to the process $pp\to j j \ell^+\ell^-\nu\bar{\nu}$.}}
\end{figure}

The process $pp\to W^+W^- jj$ can be affected by the anomalous $\gamma\gamma W W$ couplings as shown in Fig.~\ref{Fig:feynman1}.
The contribution from the tri-boson channel shown in Fig.~\ref{Fig:feynman1}.~(b) was found to be about three orders of magnitude smaller compared with the VBS contribution in Fig.~\ref{Fig:feynman1}.~(a)~\cite{aaww}, therefore in the following discussions we concentrate on the effect of the VBS contribution.
Moreover, we only consider the leptonic decays of the $W^{\pm}$ bosons, and focus on the process $pp\to \ell^+\ell^-\bar{\nu}\nu jj$ at $\sqrt{s}=13\;{\rm TeV}$, with $\ell=e,\mu$.
To distinguish, the $s$ of the subprocess $\gamma \gamma \to W^+W^-$ is denoted as $\hat{s}$.

\section{\label{level3}Approximation of the energy scale}

A prerequisite for using an ANN to mine information is that the information to be mined actually exists.
This is extremely important because the ANN is considered to be a `black box’.
To demonstrate that the $\hat{s}$ can be approximately reconstructed, and also as a comparison, we briefly introduce the method for estimating $\hat{s}$ in Ref.~\cite{aaww}.
Assuming the $W^{\pm}$ bosons are energetic and neglecting the $\mathcal{O}(M_W/\sqrt{\hat{s}})$ contributions, the leptons can be viewed as approximately collinear to the neutrinos, i.e., with $u$ and $v$ the coefficients to be determined, the momenta of the neutrinos can be related to the momenta of the charged leptons as ${\bf p}^{\nu}\approx u{\bf p}^{\ell^+}$ and ${\bf p}^{\bar{\nu}}\approx v{\bf p}^{\ell^-}$, which lead to the equations
\begin{equation}
{\bf p}^{\rm miss}_x = u{\bf p}^{\ell^+}_x+v{\bf p}^{\ell^-}_x,\;\;{\bf p}^{\rm miss}_y = u{\bf p}^{\ell^+}_y+v{\bf p}^{\ell^-}_y,
\label{eq.3.1}
\end{equation}
by which $u$ and $v$ can be solved and then $\hat{s}$ can be reconstructed. The result is
\begin{equation}
\begin{split}
&\hat{s}_{\rm ap}=\left((1+|u|)E^{\ell^+}+(1+|v|)E^{\ell^-}\right)^2\\
&-\left((1+u){\bf p}_z^{\ell^+}+(1+v){\bf p}_z^{\ell^-}\right)^2 - \left|\sum _{\pm}\mathbf{p}^{\ell^\pm}_T+\mathbf{p}^{\mathrm{miss}}_T\right|^2,
\end{split}
\label{eq.3.2}
\end{equation}
where
\begin{equation}
  u = \frac{1}{\kappa}\left({\bf p}^{\mathrm{miss}}_y {\bf p}^{\ell^-}_x -  {\bf p}^{\mathrm{miss}}_x {\bf p}^{\ell^-}_y\right), \;\;
  v = -\frac{1}{\kappa}\left({\bf p}^{\mathrm{miss}}_y {\bf p}^{\ell^+}_x - {\bf p}^{\mathrm{miss}}_x {\bf p}^{\ell^+}_y\right),
\label{eq.3.3}
\end{equation}
with
\begin{equation}
\kappa = {\bf p}^{\ell^+}_y {\bf p}^{\ell^-}_x - {\bf p}^{\ell^+}_x {\bf p}^{\ell^-}_y.
\label{eq.3.4}
\end{equation}

This approximation is based on the assumption that $W^{\pm}$ bosons are energetic, which is supported by the fact that the $\hat{s}$ are large for the signal events induced by aQGCs.
However, when $\hat{s}$ is large, the charged leptons are approximately back-to-back, and the two equations in Eq.~(\ref{eq.3.1}) will degenerate when charged leptons are exactly back-to-back.
In other words, for most events induced by aQGCs, the $\kappa $ are very small.
When $\kappa$ is close to zero, it will amplify the errors in numerator, resulting in an inaccurate approximation.

Since approximations exist, using an ANN to reconstruct $\hat{s}$ is nothing but to look for a better approximation.
The ANN is good at looking for approximations and finding patterns in complex relationships, and therefore has the potential to yield better results.

\section{\label{level4}Numerical results of the ANN}

In this section we use the ANN approach to reconstruct $\hat{s}$.
To train the ANN, we use the Monte-Carlo (MC) simulation to generate the data-sets.
We take the contributions from both diagrams of Fig.~\ref{Fig:feynman1} as the signal because they both signal the existence of the aQGCs.
As explained, the effect of Fig.~\ref{Fig:feynman1}.~(b) is negligible compared with Fig.~\ref{Fig:feynman1}.~(a).
In the following discussions, we concentrate on Fig.~\ref{Fig:feynman1}.~(a) and neglect the effect of Fig.~\ref{Fig:feynman1}.~(b), despite that the data-sets are generated with both diagrams included.

The signal events are generated by using \verb"MadGraph5_aMC@NLO"~\cite{madgraph,feynrules}, with a parton shower using \verb"Pythia82"~\cite{pythia}.
The PDF is \verb"NNPDF2.3"~\cite{NNPDF}.
A CMS-like detector simulation is applied using \verb"Delphes"~\cite{delphes}.
The events are generated assuming one operator at a time, and using the largest coefficients listed in Table~\ref{tab.1}.

The signal of the VBS process is characterized by two quark jets, events are thus required to have at least $2$ jets and two opposite sign charged leptons.
The dominant background is the process $pp\to t\bar{t}+N j$ with $t\to W^+b$~($\bar{t}\to W^+\bar{b}$) and with $b$-jet mistagged.
To reduce this background, we also require $N_j \leq 5$.
In the following, the results are established after the lepton number cut and jet number cut $N_{\ell}=2$ and $2\leq N_{j}\leq 5$.
To train the ANN, we generate $10^6$ events to build the training data-set, and another $10^6$ events to build the validation data-set for each anomalous $\gamma\gamma WW$ coupling.
After the requirement on the numbers of leptons and jets, there are about $6\times 10^5$ events in each data-set.

Before the detector simulation, the $\hat{s}$ can be obtained, which is denoted as $\hat{s}_{\rm tr}$.
Each event corresponds to an element in the data-set consists of $19$ variables.
For each event, an $18$ dimensional vector provides as the input to the ANN, which consists of $18$ variables.
They are the components of the $4$-momenta of the two hardest jets, the $4$-momenta of the two hardest opposite signed charged leptons and the components of the transverse missing momentum.
The output of the ANN corresponds to $\hat{s}$.
The true labels are the $19$-th variables of the elements in the data-sets which are $\hat{s}_{\rm tr}$ of the events.

In this section, we mainly focus on the contribution of $V_0$ vertex.
It has been found that the process $pp\to jj\ell ^+\ell ^-\nu\bar{\nu}$ is insensitive to the $V_4$ vertex~\cite{aaww}, we do not study $V_4$ in this paper.

\subsection{\label{level4.1}ANN approach}

ANN is a mathematical model to simulate the complex neural system of a human brain, and it is also an information processing system for large-scale distributed parallel information processing~\cite{ann}.
The ANN is good at finding the complex mathematical mapping relationships between input and output, it could be utilized to unveil hidden information in the final states.
The mapping relationship is determined by the number of interconnected nodes and their connection modes.
In this paper, we use a dense connected ANN.

\begin{figure}[!htbp]
\centering{
\includegraphics[width=0.7\textwidth]{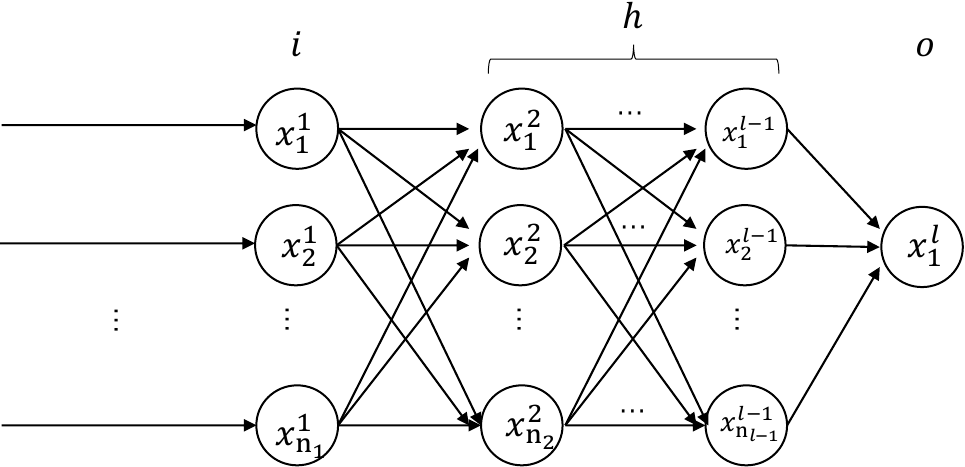}
\caption{\label{Fig:fignetwork}
The architecture of an ANN used in this paper. `i', `h' and `o' stand for input layer, hidden layer and output layer, respectively. $l$ is the number of layers and $n_i$ is the number of neurons in the $i$-th layer.}}
\end{figure}

An ANN is composed with one or more hidden layers and an output layer.
Denoting $x_j^i$ as neurons in the $i$-th layer, where $x_{1\leq j \leq n_1}^1$ are input neurons, $x_{1 \leq j \leq n_i}^{2\leq i\leq l-1}$ are in hidden layers and $x_1^l$ is the output neuron, the ANN can be depicted in Fig.~\ref{Fig:fignetwork}.

Without causing ambiguity, the value at a neuron takes the same notation $x_j^i$.
$x_{j'}^{i+1}$ can be related with $x_j^i$ as
\begin{equation}
x_{j'}^{i+1}=f^{i+1}\left(\sum _j \omega _{jj'}^{i+1}x_j^i + b_{j'}^{i+1}\right)
\label{eq.4.1}
\end{equation}
where $\omega _{jj'}^{i+1}$ are elements of a weight matrix $W^{i+1}$, $b_{j'}^{i+1}$ are components of a bias vector, and $f^{i+1}$ is an activate function.
The activation functions for the hidden layers are chosen as the parametric rectified linear unit~(PReLU) function~\cite{prelu} defined as
\begin{equation}
\begin{split}
&f(x)=\left\{\begin{array}{cc}x, & x\geq 0 ;\\ \alpha x, & x<0, \end{array}\right.\\
\end{split}
\label{eq.4.2}
\end{equation}
where $\alpha$'s are trainable parameters.
For the output layer, no activation function~(i.e., linear activation function) is used.
In this paper, without further specification we use $l=10$, $n_{10>l>1}=32$, $n_1$ is as same as the dimension of input data and $n_{10}=1$ for the output layer.
The training data-sets are normalized using the z-score standardization, i.e., denoting $v_i$ as the $i$-th variable of one of the elements in the data-sets, $v_i'$ is used instead of $v_i$ which is defined as $v'_i=(v_i-\bar{v_i})/\sigma _{v_i}$, where $\bar{v_i}$ and $\sigma _{v_i}$ are the mean value and the standard deviation of all $i$-th variables of the elements in the data-sets.

\begin{figure}[!htbp]
\centering{
\includegraphics[width=0.48\textwidth]{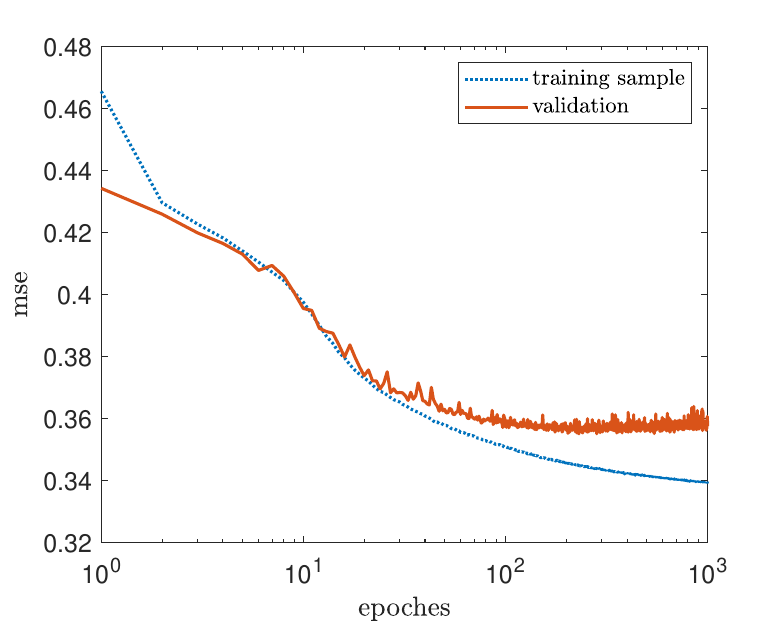}
\includegraphics[width=0.48\textwidth]{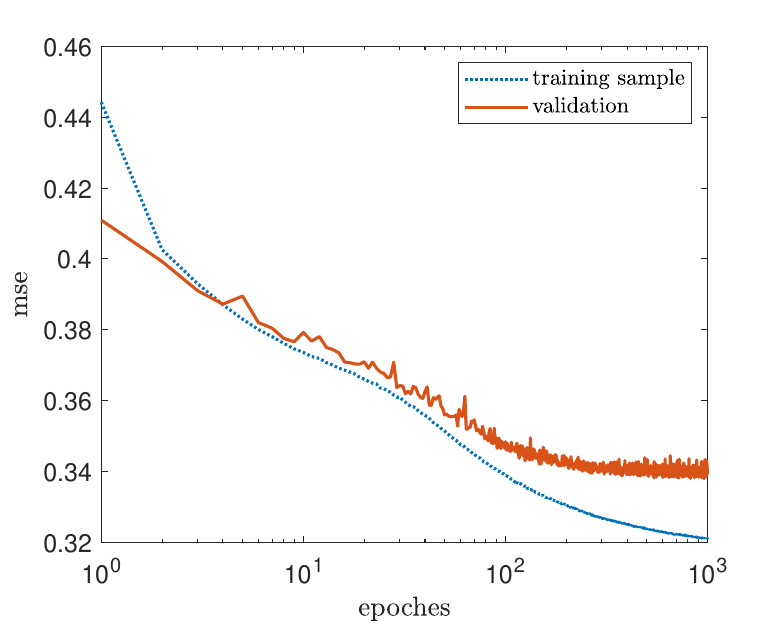}
\caption{\label{Fig:learningcurve}
The learning curves of the ANNs trained with $V_{0,2}$ data-sets.}}
\end{figure}
The architecture is built using \verb"Keras" with a \verb"TensorFlow"~\cite{tensorflow} backend.
The data preparation is performed by \verb"MLAnalysis"~\cite{Guo:2023nfu}.
The learning curves of the ANNs for $V_0$ and $V_2$ are shown in Fig.~\ref{Fig:learningcurve}.
Note that the label is also standardized, therefore the mean squared error~(mse) is dimensionless.
From Fig.~\ref{Fig:learningcurve}, we find that the mse stopped to decrease at about $150$ epoches for $V_{0,1}$ vertices, and at about $300$ epoches for $V_{2,3}$ vertices.
To avoid overfitting, we stop the training at $150$ epoches for $V_{0,1}$ vertices, and at $300$ epoches for $V_{2,3}$ vertices.
Note that the $V_{0,1}$ vertices are from $O_{M_i}$ operators and $V_{2,3}$ vertices are from the $O_{T_i}$ operators, it is interesting that the ANNs are more difficult to train with the signal events induced by $O_{T_i}$ operators.

\begin{figure}[!htbp]
\centering{
\includegraphics[width=0.7\textwidth]{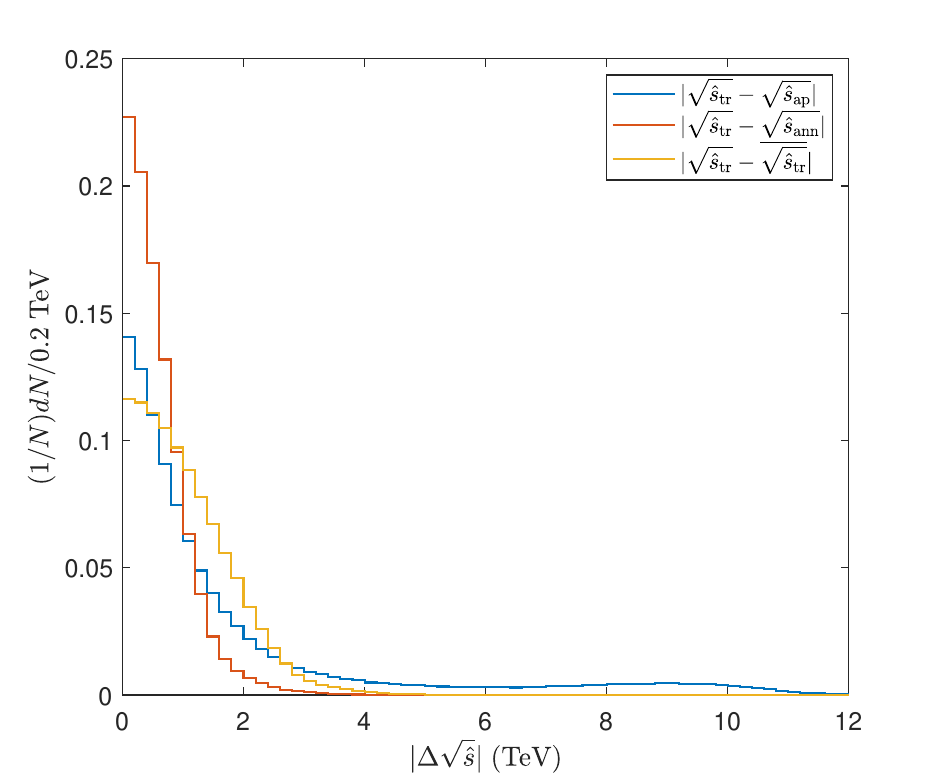}
\caption{\label{Fig:a0ann}
The normalized distributions of $|\Delta \sqrt{\hat{s}}|$ for $\overline{\sqrt{\hat{s}}}$, $\hat{s}_{\rm ap}$ and $\hat{s}_{\rm ann}$.}}
\end{figure}
Denoting $\hat{s}_{\rm ann}$ as the $\hat{s}$ predicted by the ANN.
For the validation data-set, the normalized distributions of $|\sqrt{\hat{s}_{\rm tr}}-\overline{\sqrt{\hat{s}_{\rm tr}}}|$, $|\sqrt{\hat{s}_{\rm tr}}-\sqrt{\hat{s}_{\rm ap}}|$ and $|\sqrt{\hat{s}_{\rm tr}}-\sqrt{\hat{s}_{\rm ann}}|$ are shown in Fig.~\ref{Fig:a0ann}, where $\overline{\sqrt{\hat{s}_{\rm tr}}}$ is the mean value of $\sqrt{\hat{s}_{\rm tr}}$.
One can see that the deviation of $\sqrt{\hat{s}_{\rm ap}}$ from $\sqrt{\hat{s}_{\rm tr}}$ is smaller than using $\overline{\sqrt{\hat{s}_{\rm tr}}}$ as an approximation.
On the other hand, the result of the ANN is much better than the approximation derived from the kinematical analysis.

\subsection{\label{level4.2}The information in the data-set}

It has been shown that the ANN can reconstruct $\hat{s}$ much better than the kinematical analysis.
In this subsection, we investigate how the performances of the ANNs are affected by different factors.
Specifically, we are interested in where does the information to reconstruct $\hat{s}$ contained in.
To answer this question, we pay particular attention to the information that is not used by the approximation in Eq.~(\ref{eq.3.2}).
In this subsection, the epoches are determined similarly as the previous subsection.

\subsubsection{\label{level4.2.1}Compare different sectors}

The approximation in Eq.~(\ref{eq.3.2}) does not use the momenta of jets which are difficult to be made use of.
To investigate how the results are affected by the information contained in jets, charged leptons and missing momentum, we divide the input data into 3 sectors.
We denote $\hat{s}_{\rm 2lm}$ as the $\hat{s}$ predicted by the ANN trained with the data-set consists of the components of the $4$-momenta of the two hardest opposite signed charged leptons and the transverse missing momentum,
$\hat{s}_{\rm 2jm}$ as the result of the ANN trained with the data-set consists of the components of the $4$-momenta of the two hardest jets and the transverse missing momentum,
$\hat{s}_{\rm 2j2l}$ as the result of the ANN trained with the data-set consists of the components of the $4$-momenta of the two hardest jets and the $4$-momenta of the two hardest opposite signed charged leptons,
$\hat{s}_{\rm 2j}$ as the result of the ANN trained with the data-set consists of the components of the $4$-momenta of the two hardest jets,
$\hat{s}_{\rm 2l}$ as the result of the ANN trained with the data-set consists of the components of the $4$-momenta of the two hardest opposite signed charged leptons,
$\hat{s}_{\rm m}$ as the result of the ANN trained with the data-set consists of the components of the transverse missing momentum,
respectively.

\begin{figure}[!htbp]
\centering{
\includegraphics[width=0.7\textwidth]{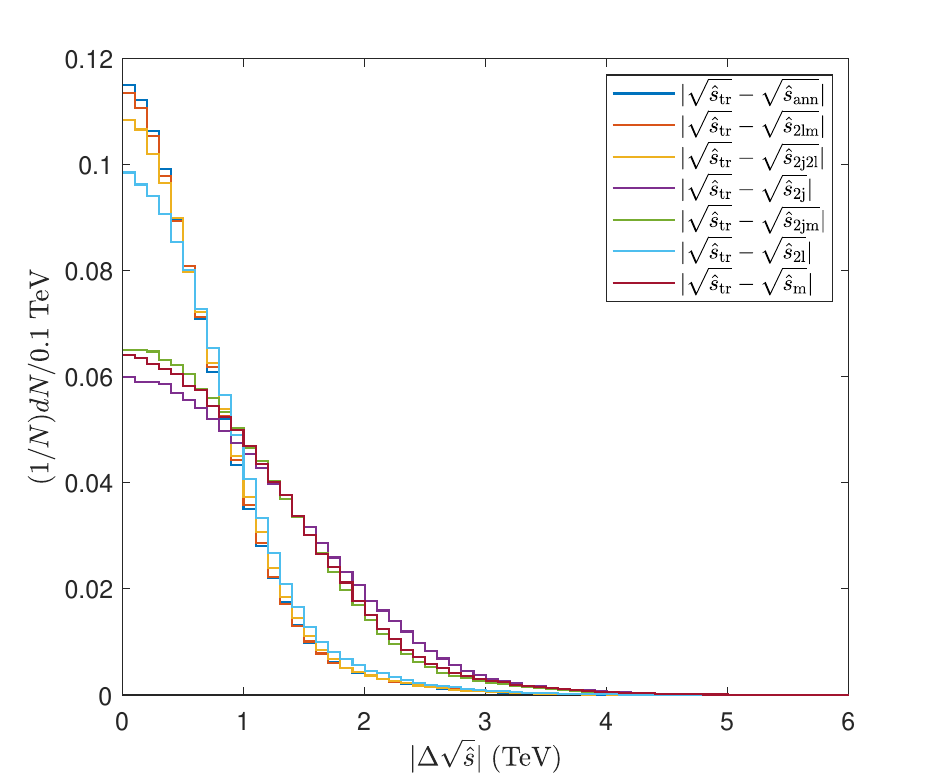}
\caption{\label{Fig:sectors}
The normalized distributions of $|\Delta \sqrt{\hat{s}}|$ for $\hat{s}_{\rm ann}$, $\hat{s}_{\rm 2lm}$, $\hat{s}_{\rm 2jm}$, $\hat{s}_{\rm 2j2l}$, $\hat{s}_{\rm 2l}$, $\hat{s}_{\rm 2j}$ and $\hat{s}_{\rm m}$.}}
\end{figure}
The normalized distributions of $|\Delta \sqrt{\hat{s}}|$ are shown in Fig.~\ref{Fig:sectors}.
From the distributions we find that the importance of different sectors can be ordered as $p^{\ell^{\pm}} > {\bf p}_T^{\rm miss}> p^{\rm jet}$.
Indeed, the ANN trained with the data-set including the momenta of jets can produce slightly more precise results.
Nevertheless, the effect brought about by jets is small and is not the main reason why ANNs are more accurate.

\subsubsection{\label{level4.2.2}Compare different operators}

Except for assuming the events are induced by aQGCs, the approximation in Eq.~(\ref{eq.3.2}) does not use any other information about the anomalous couplings.
In the approximation, the formula to estimate $\sqrt{\hat{s}}$ is same for all couplings.
Meanwhile, using the ANN approach, we can train the ANNs by using data-sets consist of signal events from different couplings.
In this way we can investigate whether the distinction between couplings is important for the reconstruction of $\sqrt{\hat{s}}$.

\begin{figure}[!htbp]
\centering{
\includegraphics[width=0.48\textwidth]{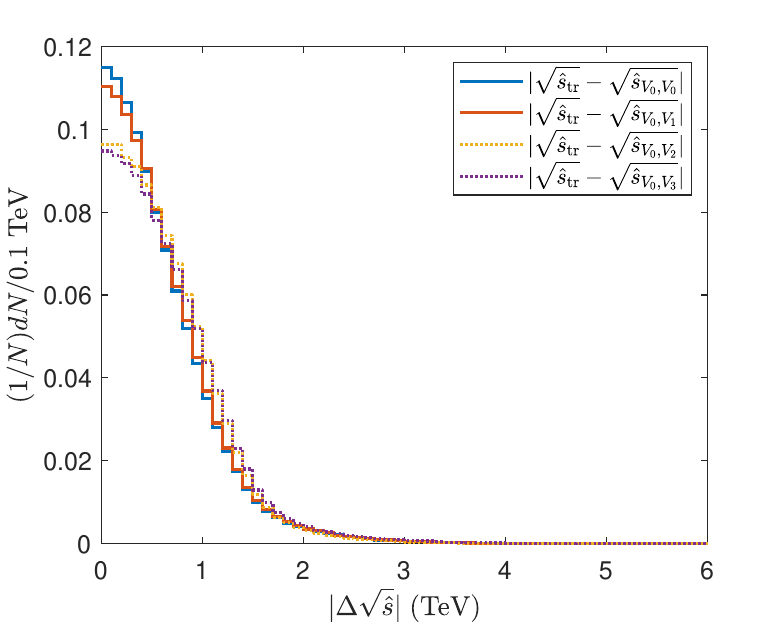}
\includegraphics[width=0.48\textwidth]{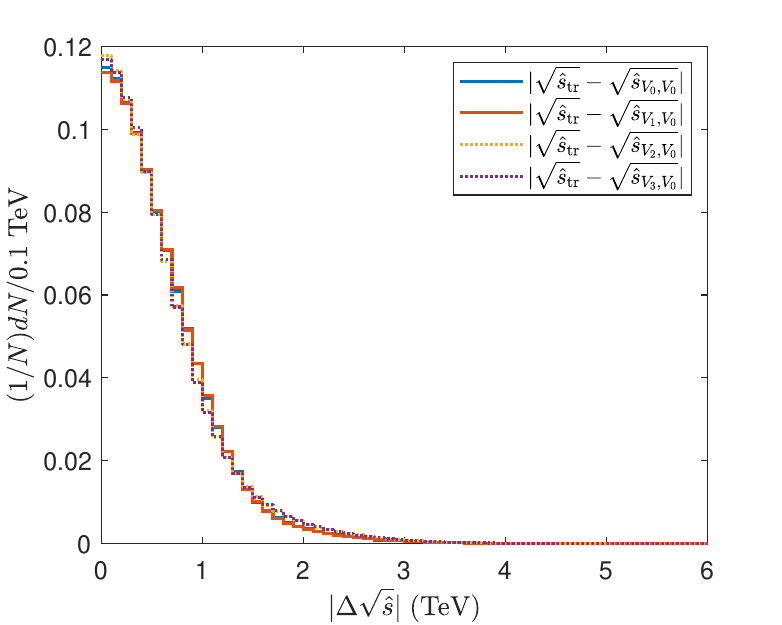}
\caption{\label{Fig:opertors}
The normalized distributions of $|\Delta \sqrt{\hat{s}}|$ for $\hat{s}_{V_0,V_i}$ and $\hat{s}_{V_i,V_0}$. Note that $\hat{s}_{V_0,V_0}$ is the $\hat{s}_{\rm ann}$ shown in Figs.~\ref{Fig:a0ann} and \ref{Fig:sectors}.}}
\end{figure}
Denoting the $\hat{s}_{V _i,V _j}$ as the predicted $\hat{s}$ of the events in a $V _j$ validation data-set but predicted by the ANN trained with the $V _i$ training data-set.
The results of $\hat{s}_{V_0,V_i}$ and $\hat{s}_{V_i,V_0}$ are shown in Fig.~\ref{Fig:opertors}.
Again, the difference between $O_{M_i}$ operators and $O_{T_i}$ operators can be found from the results of $\hat{s}_{V_0,V_i}$.
From the results of $\hat{s}_{V_i,V_0}$, it can be seen that the predictive powers of the ANNs trained using different data-sets are about the same for signal events of $V_0$ vertex.
In fact, the ANN trained with $V_3$ training data-set is slightly more accurate than the ANN trained with $V_0$ training data-set in predicting the $\hat{s}$ of the $V_0$ validation data-set.
Therefore we conclude that the information about different couplings is not made use of.
The difference in the distributions of $|\sqrt{\hat{s}_{\rm tr}}-\sqrt{\hat{s}_{V_0,V_i}}|$ is just another evidence that, from the perspectives of the ANNs the signal events induced by $O_{T_i}$ operators are more difficult to learn.

\subsubsection{\label{level4.2.3}Compare different collision energies}

\begin{figure}[!htbp]
\centering{
\includegraphics[width=0.7\textwidth]{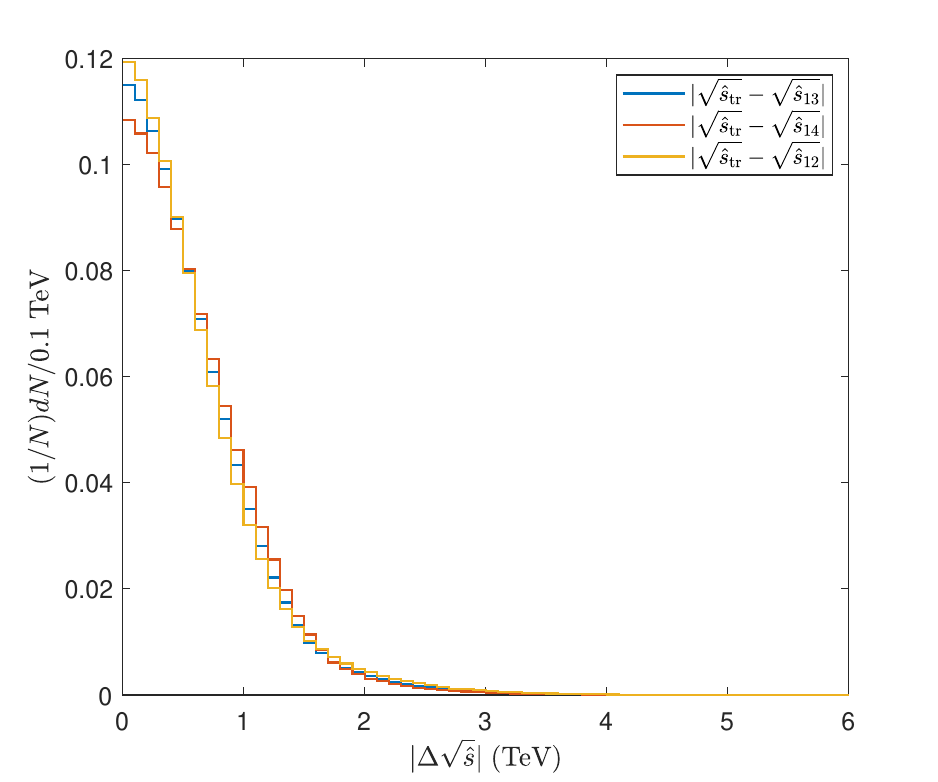}
\caption{\label{Fig:energies}
The normalized distributions of $\Delta \sqrt{\hat{s}}$ for $\hat{s}_{12,13,14}$. Note that $\hat{s}_{13}$ is the $\hat{s}_{\rm ann}$ shown in Figs.~\ref{Fig:a0ann} and \ref{Fig:sectors}.}}
\end{figure}

The approximation in Eq.~(\ref{eq.3.2}) assumes that the $\hat{s}$ is large, however it does not use the information of collision energy $\sqrt{s}=13\;{\rm TeV}$.
To investigate whether this information is made use of, we prepare three training data-sets, which are the signal events of $V_0$ generated at $\sqrt{s}=12,13$ and $14\;{\rm TeV}$.
$\hat{s}$ of the events in the $\sqrt{s}=13\;{\rm TeV}$ $V_0$ validation data-set are predicted by the ANNs trained with the $\sqrt{s}=12,13$ and $14\;{\rm TeV}$ training data-sets, which are denoted as $\hat{s}_{12,13,14}$, respectively.
The normalized distributions of $|\Delta \sqrt{\hat{s}}_{12,13,14}|$ are shown in Fig.~\ref{Fig:energies}.

It can be seen from Fig.~\ref{Fig:energies} that the results are close to each other, indicating that the $\sqrt{s}$ was also hardly used by the ANNs.
It is interesting to notice that the ANN trained with the $\sqrt{s}=12\;{\rm TeV}$ training data-set can make even better prediction on the $\sqrt{s}=13\;{\rm TeV}$ validation data-set than the ANN trained with the $\sqrt{s}=13\;{\rm TeV}$ training data-set.
In fact, the learning curve for the $\sqrt{s}=12$ TeV training data-set converges slower than the case of $13$ TeV.
We speculate that, the ANN trained with the $\sqrt{s}=12$ TeV data-set works better because the $12\;{\rm TeV}$ data-set is more difficult to train.

\section{\label{level5}Interpretation of the ANN}

To investigate how the $\sqrt{\hat{s}}$ is predicted by the ANNs, in this section, the implicity relation between $\sqrt{\hat{s}}$ and the inputs concealed in an ANN is investigated.
Since the accuracy of the ANN trained with only 4-momenta of charged leptons has been able to achieve almost the best accuracy, for simplicity, we focus on the ANN trained with only 4-momenta of the charged leptons.

Once the ANN is trained, we can input arbitrary 4-momenta of the charged leptons to study the relationship between $\hat{s}$ and the 4-momenta.
In this procedure, one can have a control on the variables, i.e., keep some variables constant and then change the others.
In contrast, using M.C. simulation for such a study is difficult because the 4-momenta of charged leptons are generated according to the probability density and therefore are not arbitrary.

One of the reasons the ANN is called a `black box' is that, although the ANN has an analytic expression, this expression is very complex and it is difficult to read the physics behind this expression.
Another motivation of this section is to find a more understandable expression.
In addition, this procedure can also be seen as a method to use the ANN to find an approximate formula.

We define $\hat{\theta}_{\ell^{\pm}}=\pi / 2 - |\beta |$ where $\beta$ is the angle between ${\bf p}^{\ell^{\pm}}$ and the ${\bf x y}$-plane, and therefore $\hat{\theta}_{\ell^{\pm}}$ is the zenith angle $\theta_{\ell^{\pm}}$ of a charged lepton when $\theta_{\ell^{\pm}}<\pi/2$ and $\pi - \theta_{\ell^{\pm}}$ when $\theta_{\ell^{\pm}}>\pi/2$, satisfying $0\leq \hat{\theta}_{\ell^{\pm}} \leq \pi / 2$.
For the $V_0$ training data-set, we find the mean values are $\bar{E}_{\ell^+}\approx \bar{E}_{\ell^+}\approx 0.915=\bar{E}$ and $\overline{\hat{\theta}}_{\ell^+}\approx \overline{\hat{\theta}}_{\ell^-}\approx 0.98=\bar{\hat{\theta}}$.

\begin{figure}[!htbp]
\centering{
\includegraphics[width=0.48\textwidth]{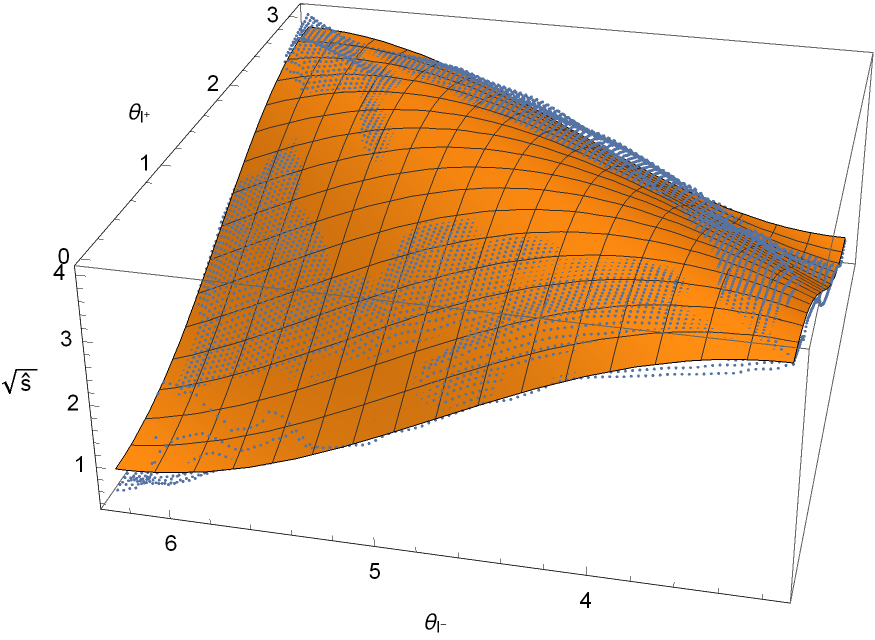}
\includegraphics[width=0.48\textwidth]{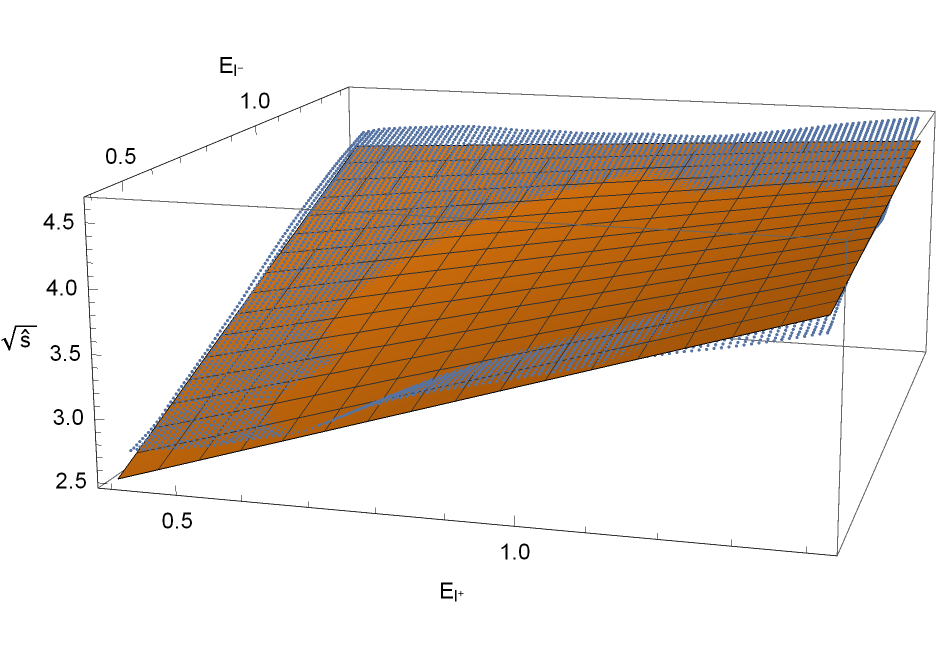}\\
\caption{\label{Fig:app1}
The relationship between $\hat{s}$ and the 4-momenta of charged leptons in Eqs.~(\ref{eq.5.1}) and (\ref{eq.5.2}) given by the ANN~(the dots). The left panel is for Eq.~(\ref{eq.5.1}) and the right panel is for Eq.~(\ref{eq.5.2}). The curved surfaces are functions fitted by using the anstaz $a_1 + a_2 \cos (\theta _{\ell ^+}-\theta _{\ell ^-})$~(the left panel) and anstaz $b_1 + b_2 (E_{\ell^+}+E_{\ell^-})+ b_3 E_{\ell^+}E_{\ell^-}$~(the right panel). }}
\end{figure}

Firstly, we use a pair of massless leptons with zero azimuth angles~(denoted as $\phi_{\ell^{\pm}}$) and with $E_{\ell^{\pm}}=\bar{E}$.
The 4-momenta are set as
\begin{equation}
\begin{split}
&p^{\ell ^+} = \bar{E}\left(1, \sin(\theta _{\ell^+}), 0, \cos(\theta _{\ell^+})\right),\\
&p^{\ell ^-} = \bar{E}\left(1, \sin(\theta _{\ell^-}), 0, \cos(\theta _{\ell^-})\right),\\
\end{split}
\label{eq.5.1}
\end{equation}
where $\theta _{\ell ^+}$ and $\theta _{\ell ^-}$ are free parameters.
The $\sqrt{\hat{s}}$ as a function of $\theta _{\ell ^+}$ and $\theta _{\ell ^-}$ predicted by the ANN is shown in the left panel of Fig.~\ref{Fig:app1}.
We find that the $\sqrt{\hat{s}}$ can be well fitted as $a_1 + a_2 \cos (\theta _{\ell ^+}-\theta _{\ell ^-})$, which is shown as the curved surface.

Moreover, we also investigate how $\sqrt{\hat{s}}$ depends on $E_{\ell^+}$ and $E_{\ell^-}$.
For this purpose, we introduce a pair of back-to-back massless leptons with the directions of ${\bf p}^{\ell ^+}$ and ${\bf p}^{\ell ^-}$ fixed, i.e., we use the following 4-momenta
\begin{equation}
\begin{split}
&p^{\ell ^+} = E_{\ell^+}\left(1, \sin(\bar{\hat{\theta}}), 0, \cos(\bar{\hat{\theta}})\right),\\
&p^{\ell ^-} = E_{\ell^-}\left(1, -\sin(\bar{\hat{\theta}}), 0, -\cos(\bar{\hat{\theta}})\right),\\
\end{split}
\label{eq.5.2}
\end{equation}
where $E_{\ell^+}$ and $E_{\ell^-}$ are free parameters.
The $\sqrt{\hat{s}}$ as a function of $E_{\ell^+}$ and $E_{\ell^-}$ predicted by the ANN is shown in the right panel of Fig.~\ref{Fig:app1}.
We find that the $\sqrt{\hat{s}}$ can be well fitted as $b_1 + b_2 (E_{\ell^+}+E_{\ell^-})+ b_3 E_{\ell^+}E_{\ell^-}$, which is shown as the curved surface.

\begin{figure}[!htbp]
\centering{
\includegraphics[width=0.35\textwidth]{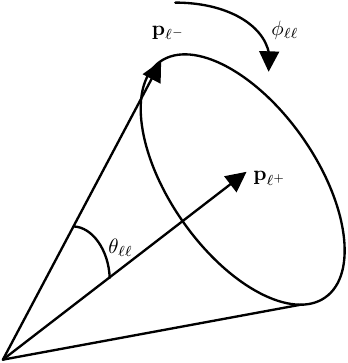}
\includegraphics[width=0.6\textwidth]{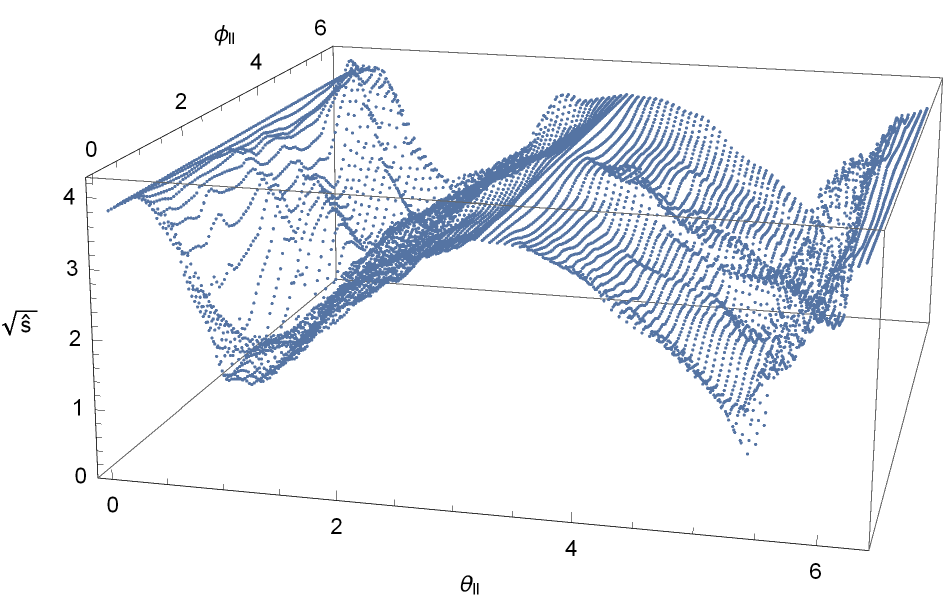}\\
\caption{\label{Fig:app2}
The directions of the momenta of charged leptons used to study the relationship between $\hat{s}$ and the angles between charged leptons are shown in the left panel. The $\hat{s}$ as a function of $\theta_{\ell\ell}$ and $\phi_{\ell\ell}$ given by the ANN is shown in the right panel.}}
\end{figure}

Denoting the angle between the momenta of $\ell^{\pm}$ as $\theta _{\ell\ell}$, the relation between $\sqrt{\hat{s}}$ and $\theta _{\ell\ell}$ is investigated.
In this case, we use $p^{\ell^+}=\bar{E}\left(1, \sin(\bar{\hat{\theta}}), 0, \cos(\bar{\hat{\theta}})\right)$, and let ${\bf p}_{\ell^-}$ be on the surface of a cone whose central axis is ${\bf p}^{\ell^+}$, with $E_{\ell^-}=\bar{E}$.
The ${\bf p}^{\ell^+}$ and ${\bf p}^{\ell^-}$ are depicted in the left panel of Fig.~\ref{Fig:app2} where the definition of $\phi _{\ell\ell}$ is shown.
The $\sqrt{\hat{s}}$ as a function of $\theta_{\ell\ell}$ and $\phi_{\ell\ell}$ given by the ANN is shown in the right panel of Fig.~\ref{Fig:app2}.
We find that, $\sqrt{\hat{s}}$ is insensitive to $\phi _{\ell\ell}$.
For the case of aQGCs, the $W^{\pm}$ bosons are typically energetic.
As a result, the charged leptons are dominantly back-to-back.
We find that, at the vicinity of $\theta _{\ell\ell}\approx \pi$, $\sqrt{\hat{s}}$ is almost independent of $\phi _{\ell\ell}$, and is approximately a cosine function of $\theta _{\ell\ell}$.

Based on the observations, we assume the relation between $\sqrt{\hat{s}}$ and the momenta of charged leptons can be fitted by an ansatz in Eq.~(\ref{eq.5.3}).
Note that, for fixed $E_{\ell^{\pm}}$ and $\phi_{\ell^{\pm}}=0$, the ansatz has the form $a_1+a_2\cos (\theta _{\ell^+}-\theta _{\ell^-})$.
For fixed $\theta_{\ell^{\pm}}$ and $\phi_{\ell^{\pm}}=0$, the ansatz has the form $b_1 + b_2 (E_{\ell^+}+E_{\ell^-})+ b_3 E_{\ell^+}E_{\ell^-}$.
Besides, the ansatz is a cosine function of $\theta _{\ell\ell}$ and is independent of $\phi _{\ell\ell}$.
\begin{equation}
\begin{split}
&\sqrt{\hat{s}}\approx c_1 + c_2 \left(E_{\ell^+}+E_{\ell^-}\right)+\left(c_3 + c_4 \left(E_{\ell^+}+E_{\ell^-}\right)+c_5 E_{\ell^+} E_{\ell^-}\right)\cos (\theta _{\ell\ell})\\
&c_1 = 1.33642\;{\rm TeV}, \;\;c_2 = 0.511962,\\
&c_3 = -0.355933\;{\rm TeV},\;\; c_4 = -0.778735,\;\; c_5 = 0.298186\;{\rm TeV}^{-1}
\end{split}
\label{eq.5.3}
\end{equation}

Using the training data-set, the relation between $\sqrt{\hat{s}}$ and the momenta of charged leptons are fitted using the ansatz in Eq.~(\ref{eq.5.3}), the results are also shown in Eq.~(\ref{eq.5.3}).

\begin{figure}[!htbp]
\centering{
\includegraphics[width=0.7\textwidth]{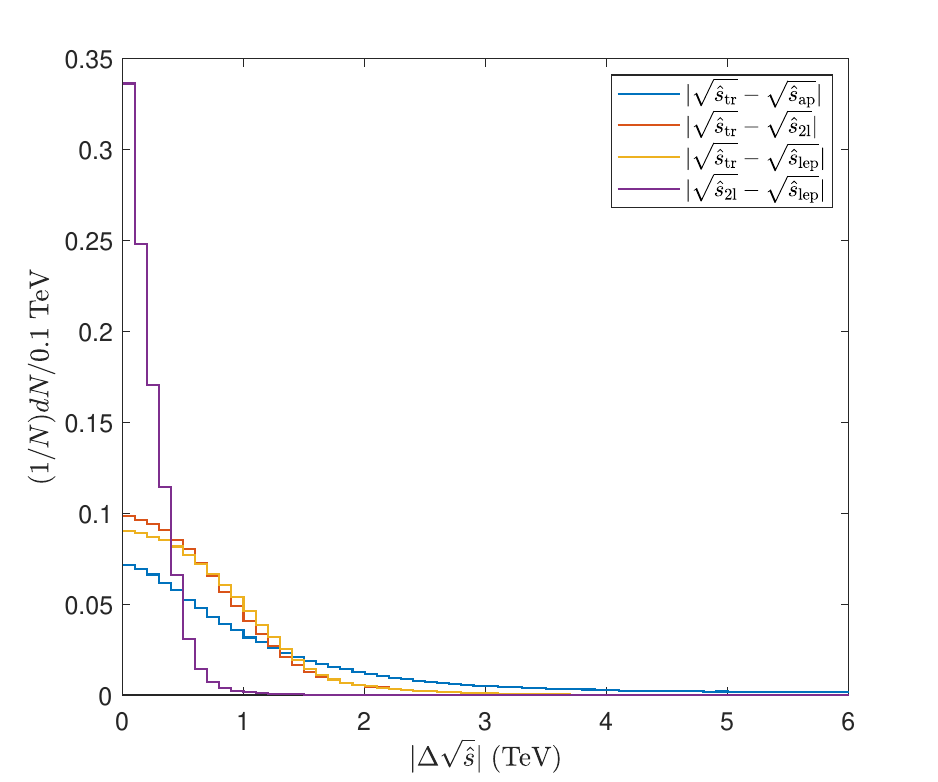}
\caption{\label{Fig:lepapp}
The normalized distributions of $|\sqrt{\hat{s}_{\rm lep}}-\sqrt{\hat{s}_{\rm 2l}}|$, $|\sqrt{\hat{s}_{\rm tr}}-\sqrt{\hat{s}_{\rm lep}}|$, $|\sqrt{\hat{s}_{\rm tr}}-\sqrt{\hat{s}_{\rm 2l}}|$ and $|\sqrt{\hat{s}_{\rm tr}}-\sqrt{\hat{s}_{\rm ap}}|$.}}
\end{figure}

Denoting $\sqrt{\hat{s}_{\rm lep}}$ as the $\sqrt{\hat{s}}$ predicted by Eq.~(\ref{eq.5.3}), the normalized distribution of $|\sqrt{\hat{s}_{\rm lep}}-\sqrt{\hat{s}_{\rm 2l}}|$ for the $V_0$ validation data-set is shown in Fig.~\ref{Fig:lepapp}.
We find that, Eq.~(\ref{eq.5.3}) can approximate the ANN well.
The normalized distributions of $\Delta \sqrt{\hat{s}}$ for $\hat{s}_{\rm ap}$, $\hat{s}_{\rm lep}$ and $\hat{s}_{\rm 2l}$ are also shown in Fig.~\ref{Fig:lepapp}.
It can be seen that, Eq.~(\ref{eq.5.3}) is able to achieve comparable results to the ANN.
Meanwhile, as an approximation found by the ANN, Eq.~(\ref{eq.5.3}) is much better than Eq.~(\ref{eq.3.2}).

The ANN with $n_1=8$ contains $9025$ trainable parameters.
As a contrast, the ansatz in Eq.~(\ref{eq.5.3}) has only five fitting parameters.
Besides, Eq.~(\ref{eq.5.3}) is no longer an overly complicated expression which is hard to read.
One can already see some patterns from Eq.~(\ref{eq.5.3}).
For example, $\hat{s}$ is insensitive to $\phi _{\ell\ell}$, and is approximately a linear function of $E_{\ell^+}+E_{\ell^-}$ when $\cos (\theta _{\ell\ell})=0$.

\section{\label{level6}Expected constraints on the aQGCs}

Where to find the signals of the NP is one of the most important questions in the study of NP.
Since the signals of aQGCs are not observed yet, the objective of this section is to investigate the sensitivity of VBS processes to the aQGCs.
In this section we study the signals and backgrounds with the help of the ANN.
To take into account the unitarity bounds, $\hat{s}$ is necessary.
The $\hat{s}$ reconstructed by the ANN approach is made use of to apply the unitarity bounds which are important in the study of an EFT.

\subsection{\label{level6.1}One ANN for all couplings}

We have confirmed in Sec.~\ref{level4} that the ANN does not use the information about which coupling the events come from.
For simplicity, and on the other hand, to have more sufficient training samples, we combine the training data-sets of $V_{0,1,2,3}$ to one data-set, and use this data-set to train one ANN for all couplings.

\begin{figure}[!htbp]
\centering{
\includegraphics[width=0.48\textwidth]{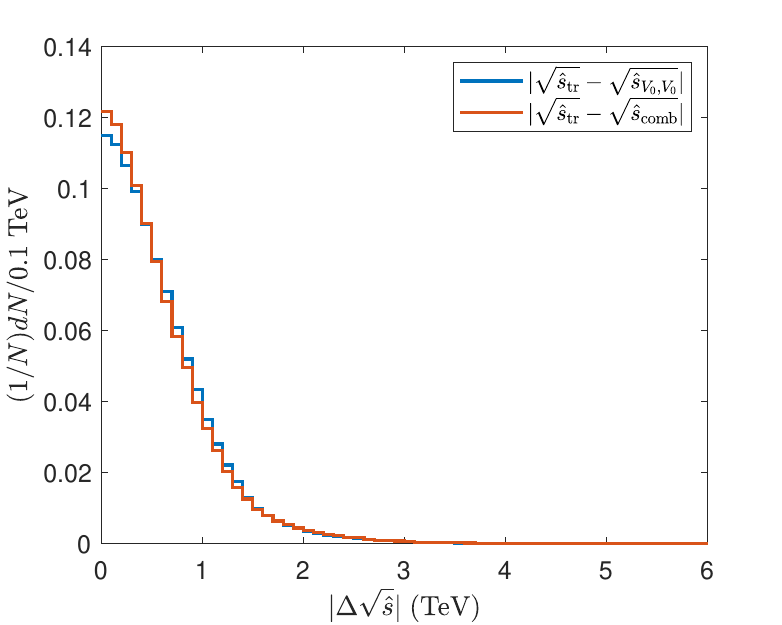}
\includegraphics[width=0.48\textwidth]{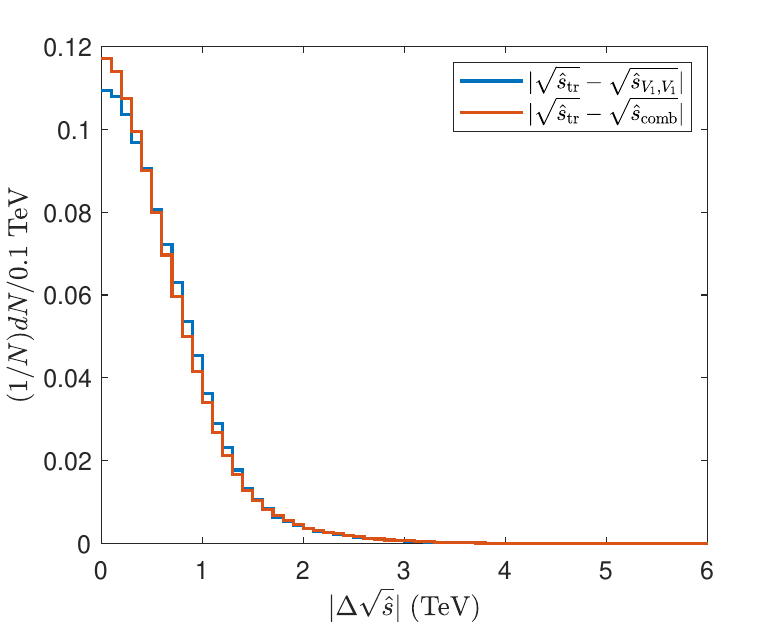}\\
\includegraphics[width=0.48\textwidth]{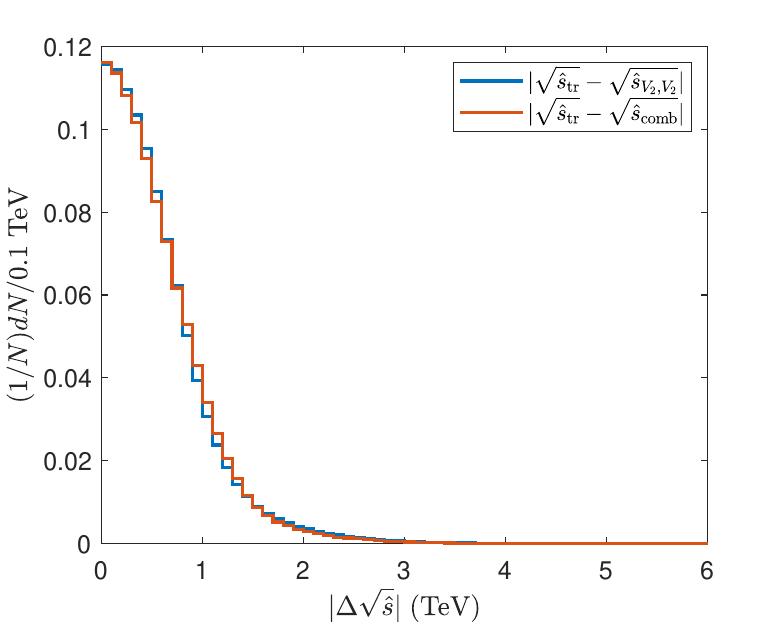}
\includegraphics[width=0.48\textwidth]{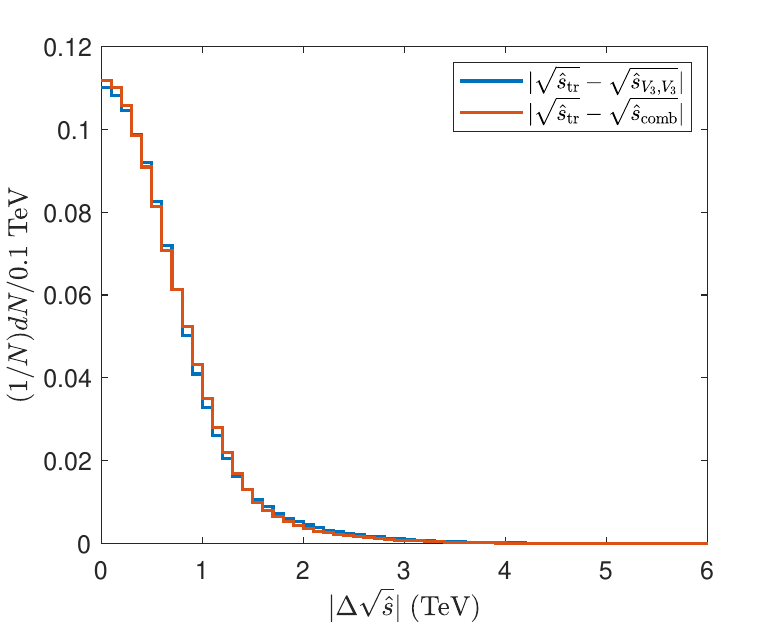}\\
\caption{\label{Fig:finalAnn}
The normalized distributions of $\Delta \sqrt{s}$ for $\hat{s}_{\rm comb}$ compared with those for $\hat{s}_{V_i,V_i}$.}}
\end{figure}
Denoting $\hat{s}_{\rm comb}$ as $\hat{s}$ of the events in validation data-sets predicted by the ANN trained with the combined data-set, normalized distributions of $\Delta \sqrt{\hat{s}}$ for $\hat{s}_{\rm comb}$ and $\hat{s}_{V_i,V_i}$ are compared in Fig.~\ref{Fig:finalAnn}.
We find that for $V_{0,1}$ vertices, $\hat{s}_{\rm comb}$ are slightly better than $\hat{s}_{V_0,V_0}$ and $\hat{s}_{V_1,V_1}$, for $V_{2,3}$ vertices, $\hat{s}_{\rm comb}$ are about the same as $\hat{s}_{V_2,V_2}$ and $\hat{s}_{V_3,V_3}$.
In the remainder of this section, we use $\hat{s}_{\rm comb}$.

\subsection{\label{level6.2}Signals and backgrounds}

\begin{figure}[!htbp]
\centering{
\includegraphics[width=0.7\textwidth]{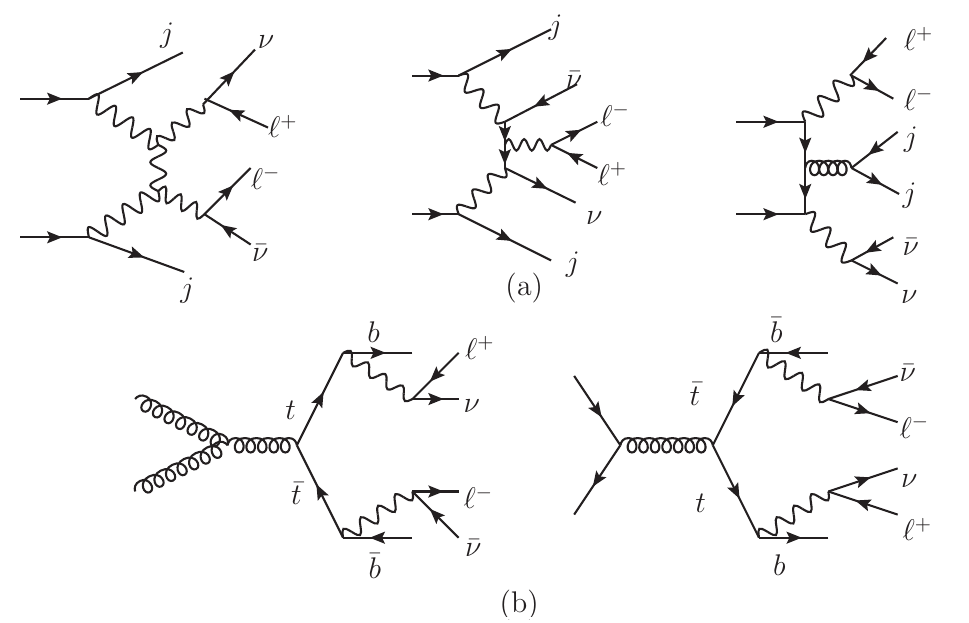}
\caption{\label{Fig:background}
The typical Feynman diagrams of the backgrounds.}}
\end{figure}
At the LHC, the $t\bar{t}+N j$ production contribute to the backgrounds due to the $b$-jet mistag.
The Feynman diagrams in the case of $N=0$ are shown in Fig.~\ref{Fig:background}.~(b).
The cross section of inclusive $t\bar{t}$ production is about $888$ pb~\cite{ttbarcs}, with the $77\%$~\cite{btag} $b$-tagging efficiency, the inclusive $t\bar{t}$ production would lead to a $jj\ell^+\ell^-\nu\bar{\nu}$ background whose cross section is about $2.32$ pb.
Apart from the $t\bar{t}$ backgrounds, significant irreducible backgrounds can arise from the SM processes which lead to the same final state $\ell^+ \ell^- \nu \bar{\nu} j j$.
The typical Feynman diagrams at tree level are shown in Fig.~\ref{Fig:background}.~(a), which are often categorized as the electroweak VBS~(EW-VBS), electroweak non-VBS~(EW-non-VBS) and QCD processes.
To highlight the contributions from aQGCs, the contributions from EW-VBS diagrams including those contain the SM $\gamma\gamma W^+W^-$ coupling are also considered as parts of backgrounds.
In the following, the backgrounds shown in Fig.~\ref{Fig:background}.~(a) are denoted as `SM', and the backgrounds in Fig.~\ref{Fig:background}.~(b) are denoted as `$t\bar{t}$'.

\begin{figure}[!htbp]
\centering{
\includegraphics[width=0.7\textwidth]{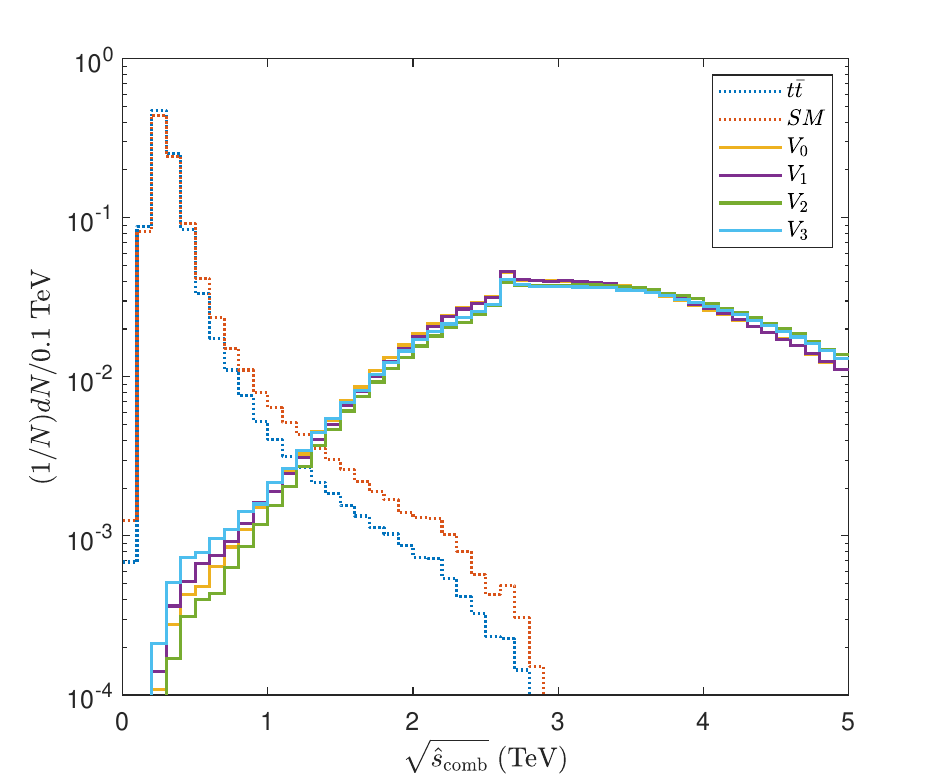}
\caption{\label{Fig:shatdist}
The normalized distributions of $\sqrt{\hat{s}_{\rm comb}}$ for the signal events and background events.}}
\end{figure}

We use the event selection strategy in Ref.~\cite{aaww},
\begin{equation}
\begin{split}
&M_{jj}>150{\rm\ GeV},\ \Delta y_{jj}>1.2,\;\;|\cos (\phi _{LM})| > 0.3,\\
&\cos(\theta _{\ell\ell})< 0,\;\;\hat{s}_{\rm ap} > 1.5 \;{\rm TeV}^2,\;\;M_{o1}>600{\rm\ GeV},
\end{split}
\label{eq.6.1}
\end{equation}
where $M_{jj}$ and $\Delta y_{jj}$ are invariant mass and difference between the rapidities of the hardest two jets, $\phi _{LM}$ is the angle between the transverse missing momentum and the sum of transverse momenta of charged leptons, i.e. the angle between ${\bf p}^{\ell^+}_T+{\bf p}^{\ell^-}_T$ and ${\bf p}^{\rm miss}_T$, $\theta _{\ell\ell}$ is the angle between the charged leptons, and~\cite{mo1}
\begin{equation}
\begin{split}
&M_{o1}\equiv \sqrt{\left(|{\bf p}_T^{\ell^+}|+|{\bf p}_T^{\ell^-}|+|{\bf p}^{\rm miss}_{\rm T}|\right)^2-\left|{\bf p}_T^{\ell^+}+{\bf p}_T^{\ell^-}+{\bf p}^{\rm miss}_{\rm T}\right|^2},\\
\end{split}
\label{eq.6.2}
\end{equation}
which was found to be very efficient to highlight the signals of aQGCs in the study of same sign WW scattering.
In this paper, we use $\hat{s}_{\rm comb}$ instead of $\hat{s}_{\rm ap}$ and adjust the cut to $\hat{s}_{\rm comb}>2.25\;{\rm TeV}^2$.
$\hat{s}$ is not well defined in the cases such as the $t\bar{t}$ background events, nevertheless, $\hat{s}_{\rm comb}$ can still serve as an observable to discriminate the signal events from the backgrounds.
The normalized distributions of $\sqrt{\hat{s}_{\rm comb}}$ are shown in Fig.~\ref{Fig:shatdist}.
It can be seen that, for a background event, $\hat{s}_{\rm comb}$ is generally smaller than $2.25\;{\rm TeV}^2$, which is not the case for a signal event.

By using the event selection strategy in Eq.~(\ref{eq.6.1}), the cut flow is shown in Table~\ref{tab.cuts}.
\begin{table}[t]
\begin{center}
\begin{tabular}{c|c|c|c|c|c|c}
\hline
$\sigma$($\rm fb$) &$\alpha _0=0.013$ & $\alpha _1=0.021 $ & $\alpha _2=0.38 $ & $\alpha _3=0.69$  &  SM & $t\bar{t}$\\
&(TeV$^{-2}$)&(TeV$^{-2}$)&(TeV$^{-4}$)&(TeV$^{-4}$)& \\
\hline
before cuts                                    &$1.66$  &$0.29$  &$0.49$  &$0.12$  & $724.16$ & $24938.24$ \\
\hline
$N_\ell=2$                                     &$0.91$  &$0.16$  &$0.26$  &$0.064$ & $308.63$ & $8162.41$ \\
$5\geq N_j\geq 2$ & & & & & \\
\hline
$M_{jj}>150{\rm\ GeV}$                         &$0.76$  &$0.13$  &$0.22$  &$0.053$ & $111.44$  &$2766.76$ \\
$\Delta y_{jj}>1.2$  & & & & & \\
\hline
$|\cos (\phi _{L m})|>0.3$                     &$0.69$  &$0.12$  &$0.21$  &$0.050$ & $91.13$  &$2233.19$ \\
\hline
$\cos (\theta _{\ell\ell})<0$                  &$0.69$  &$0.12$  &$0.20$  &$0.049$ & $49.09$  &$1301.81$ \\
\hline
$M_{o1}>600\;{\rm GeV}$                        &$0.64$  &$0.11$  &$0.20$  &$0.046$ & $0.61$  &$2.34$ \\
\hline
$\hat{s}_{\rm comb}>2.25\;{\rm TeV}^2$         &$0.64$  &$0.11$  &$0.19$  &$0.046$ & $0.38$  &$1.60$ \\
\hline
\end{tabular}
\end{center}
\caption{\label{tab.cuts} The effect of the cuts on the process $pp\to\ell^+\ell^-\nu\bar{\nu}jj$. The cross sections of signal and background events are given in fb. The result of $t\bar{t}+Nj$ with $N=0$ is shown as an example, the effect of $b$-tagging is not included in this table which will reduce the cross-section from $24938.24\;{\rm fb}$ to $1319.23\;{\rm fb}$ with $77\%$ $b$-tagging efficiency.}
\end{table}
From Table~\ref{tab.cuts}, it can be seen that the cuts can reduce the backgrounds significantly while preserving much of the signals events.
For the $pp\to t\bar{t}+Nj$ backgrounds, the cut efficiency for $N = 0$ is the lowest mainly because the $N_j\leq 5$ cut can increase the cut efficiencies for the cases of $N>0$.
An upper bound of the $pp\to t\bar{t}+Nj$ backgrounds is estimated by using the efficiency of $N=0$ for all values of N.
Then, the $pp\to t\bar{t}+Nj$ backgrounds is reduced from $2.32$ (pb) to about $0.15$ (fb).

\subsection{\label{level6.3}Unitarity bounds}

As an EFT, the SMEFT is only valid under a certain energy scale.
The cross-section of the VBS process with contributions from aQGCs included grows significantly at high energies.
On one hand, at higher energies the VBS process is ideal to search for aQGCs.
On the other hand, the cross-section will violate unitarity at a certain high energy, which provides a signature indicating that the SMEFT is not valid.
The violation of unitarity can be avoided by unitarization methods such as K-matrix unitarization~\cite{vbfcut}, T-matrix unitarization~\cite{jrr2}, form factor method~\cite{wwexp1,vbfcut}, as well as dispersion relation method~\cite{dispersion1,dispersion2}.
It has been pointed out that, the constraints on the coefficients dependent on the method used~\cite{unitarizationeffects}, and it has been emphasised that unitrization defeats the model-independent purpose of using an EFT~\cite{vbsreview}.
Therefore, we present our results using a procedure independent of unitarization methods.

Considering the subprocess $\gamma_{\lambda _1}\gamma_{\lambda _2}\to W^-_{\lambda _3}W^+_{\lambda _4}$, where $\lambda _{1,2}=\pm 1$ and $\lambda _{3,4}=\pm 1, 0$ correspond to the helicities of the vector bosons, in the c.m. frame of two photons with ${\bf z}$-axis along the flight direction of $\gamma _{\lambda _1}$, the amplitudes can be expanded as~\cite{partialwaveexpansion}
\begin{equation}
\begin{split}
&\mathcal{M}(\gamma_{\lambda _1}\gamma_{\lambda _2}\to W^-_{\lambda _3}W^+_{\lambda _4})=8\pi \sum _{J}\left(2J+1\right)\sqrt{1+\delta _{\lambda _1\lambda _2}}e^{i(\lambda-\lambda ') \phi}d^J_{\lambda \lambda '}(\theta) T^J\\
\end{split}
\label{eq.6.3}
\end{equation}
where $\theta$ and $\phi$ are zenith and azimuth angles of the $W^-$ boson, $\lambda = \lambda _1-\lambda _2$, $\lambda ' =\lambda _3-\lambda _4$ and $d^J_{\lambda \lambda '}(\theta)$ are the Wigner D-functions~\cite{partialwaveexpansion}.
The partial wave unitarity bound is $|T^J|\leq 2$~\cite{partialwaveunitaritybound}.

For the $\gamma \gamma \to W ^+W^-$, $36$ different helicity amplitudes can be obtained.
The number of amplitudes can be reduced by using $\mathcal{M}_{\lambda _1, \lambda _2, \lambda _3, \lambda _4}(\theta ) = (-1)^{\lambda _1-\lambda _2-\lambda _3+\lambda _4}\mathcal{M}_{-\lambda _1, -\lambda _2, -\lambda _3, -\lambda _4}(\theta )$.
It is only necessary to keep the terms at the leading order~($\mathcal{O}(\hat{s}^2)$).
The helicity amplitudes at the leading order are list in Table.~\ref{tab.Helicity}.
\begin{table}
\begin{center}
\begin{tabular}{c|ccccc}
\hline
Amplitudes & $\alpha _0$ & $\alpha _1$ & $\alpha _2$ & $\alpha _3$ & $\alpha _4$ \\
\hline
$\mathcal{M}_{++++}$ & $\mathcal{O}(\hat{s})$ & $\mathcal{O}(\hat{s})$ & $2\alpha _2 \hat{s}^2$ & $\frac{1}{2}\alpha _3 \hat{s}^2$ & $\mathcal{O}(\hat{s}^0)$ \\
$\mathcal{M}_{++--}$ & $\mathcal{O}(\hat{s})$ & $\mathcal{O}(\hat{s})$ & $2\alpha _2 \hat{s}^2$ & $\frac{1}{2}\alpha _3 \hat{s}^2$ & $\frac{1}{4}\alpha _4 \hat{s}^2 \left(\cos(2\theta)+3\right)$ \\
$\mathcal{M}_{++00}$ & $\alpha _0\frac{\hat{s}^2}{M_W^2}$ & $\frac{1}{4}\alpha _1\frac{\hat{s}^2}{M_W^2}$ & $\mathcal{O}(\hat{s})$ & $\mathcal{O}(\hat{s})$ & $\mathcal{O}(\hat{s})$ \\
$\mathcal{M}_{+-+-}$ & $\mathcal{O}(\hat{s}^0)$ & $\mathcal{O}(\hat{s})$ & $\mathcal{O}(\hat{s}^0)$ & $\frac{1}{2}e^{2i\phi}\alpha _3 \hat{s}^2\cos^4(\frac{\theta}{2})$ & $\alpha _4 e^{2i\phi}s^2\cos^4(\frac{\theta}{2})$ \\
$\mathcal{M}_{+--+}$ & $\mathcal{O}(\hat{s}^0)$ & $\mathcal{O}(\hat{s})$ & $\mathcal{O}(\hat{s}^0)$ & $\frac{1}{2}e^{-2i\phi}\alpha _3\hat{s}^2\sin^4(\frac{\theta}{2})$ & $\alpha _4e^{-2i\phi}\hat{s}^2\sin^4(\frac{\theta}{2})$ \\
$\mathcal{M}_{+-00}$ & $\mathcal{O}(\hat{s}^0)$ & $\frac{e^{2i\phi}\alpha _1 \hat{s}^2 \sin ^2(\theta)}{8 M_W^2}$ & $\mathcal{O}(\hat{s}^0)$ & $\mathcal{O}(\hat{s})$ & $\mathcal{O}(\hat{s})$ \\
\hline
\end{tabular}
\end{center}
\caption{\label{tab.Helicity}The helicity amplitudes at the order of $\mathcal{O}(s^2)$.}
\end{table}
The tightest bounds are
\begin{equation}
\begin{split}
&\hat{s}^2\leq \frac{16\sqrt{2}\pi M_W^2}{\left|\alpha _0\right|},\;\;\hat{s}^2\leq \frac{64\sqrt{2}\pi M_W^2}{\left|\alpha _1\right|},\;\;
 \hat{s}^2\leq \frac{8\sqrt{2}\pi}{\left|\alpha _2\right|},\;\;\hat{s}^2\leq \frac{32\sqrt{2}\pi }{\left|\alpha _3\right|},\;\;\hat{s}^2\leq \frac{24\sqrt{2}\pi}{\left|\alpha _4\right|}.
\end{split}
\label{eq.6.4}
\end{equation}

The partial wave unitarity bound has been widely used in previous studies~\cite{unitarity1,unitarity2,unitarity3,ssww,ubnew1,ubnew2,ubnew3,ntgc5}.
To avoid the violation of unitarity, the partial wave unitarity bound was often used as constraints on the coefficients of the high dimensional operators.
Note that, in Eq.~(\ref{eq.6.4}), the unitarity bounds are presented as constraints on $\hat{s}$ instead of the coefficients.
Due to the PDF, the $\hat{s}$ of the subprocess is not a fixed value, which brings difficulties in setting constraints on the coefficients directly.
Therefore, in this paper we use a matching procedure~\cite{matchingidea2,matchingidea3} instead.
The matching procedure is built based on the idea that, to take validity into account, the constraints obtained by experiments should be reported as functions of energy scales~\cite{matchingidea1}, and has been introduced in the studies of the aQGCs~\cite{wastudy,za}.
Such a matching procedure is independent of unitarization methods and can be applied in experiments.
The matching procedure in this paper is also very similar to the `clipping' method which also cuts off the signal events violating unitarity according to $\hat{s}$~\cite{eventclipping1,eventclipping2}, except that we also cut off the backgrounds so one can compare the signals with backgrounds under a same $\hat{s}$ cut.

We use Eq.~(\ref{eq.6.4}) as a cut on $\hat{s}$, and compare the cross-sections with and without aQGCs under a same energy cut.
We shall emphasis that, although this approach is called `unitarity bound', using this approach we are actually not applying any constraints or unitarizations.
In any case, it is practicable to compare NP and the SM under a certain energy scale.
Especially, it is necessary in the detailed study of the Wilson coefficients in an EFT because the Wilson coefficients are functions of the energy scale.
This matching procedure is independent of whether or not the unitarity bounds are imposed.
We merely choose a matching energy scale such that the unitarity is guaranteed.
Specifically, we choose the energy scale as the maximally allowed energy scale according to the coefficients of the aQGCs in the sense of unitarity.

\begin{table}[!htbp]
\begin{center}
\begin{tabular}{c|c|c|c|c}
\hline
$\sigma$($\rm fb$) &$\alpha _0=0.013$ & $\alpha _1=0.021 $ & $\alpha _2=0.38 $ & $\alpha _3=0.69$ \\
&(TeV$^{-2}$)&(TeV$^{-2}$)&(TeV$^{-4}$)&(TeV$^{-4}$) \\
\hline
$\sqrt{\hat{s}}_{\rm max}\;({\rm TeV})$             &$2.44$  &$3.06$  &$3.11$  &$2.79$   \\
\hline
before unitarity bounds~(fb)                      &$0.64$  &$0.11$  &$0.19$  &$0.046$   \\
after unitarity bounds~(fb)                       &$0.083$  &$0.042$  &$0.068$  &$0.028$   \\
\hline
\end{tabular}
\end{center}
\caption{\label{tab.unitaritybounds}The $\sqrt{\hat{s}}_{\rm max}$ correspond to the largest coefficients in the ranges listed in Table~\ref{tab.1}, and the cross-sections before and after the energy cuts in Eq.~(\ref{eq.6.4}).}
\end{table}
For the largest coefficients listed in Table~\ref{tab.1}, the maximally allowed energy scales~(denoted as $\sqrt{\hat{s}}_{\rm max}$) according to Eq.~(\ref{eq.6.4}) are listed in Table~\ref{tab.unitaritybounds}.
The effect of the unitarity bounds are also shown in Table~\ref{tab.unitaritybounds}.
It can be seen that, the unitarity bounds have great suppressive effects on the cross-sections.
Especially for $V_0$, the cross-section is reduced by about an order of magnitude.
Such a significant suppression indicates the necessity of the unitarity bounds.

\subsection{\label{level6.4}Signal significance}

The sensitivity of the process $pp\to jj \ell^+\ell^-\nu\bar{\nu}$ to the aQGCs can be estimated with the help of statistical significance defined as $\mathcal{S}_{stat}\equiv N_S/\sqrt{N_S+N_B}$, where $N_S$ is the number of signal events, and $N_B$ is the number of the background events.
It has been shown that, within the current ranges of coefficients of the aQGCs, the interference terms can also be neglected~\cite{aaww}.
For simplicity, the cross-sections are calculated with the above contributions neglected.

\begin{figure}[!htbp]
\centering{
\includegraphics[width=0.48\textwidth]{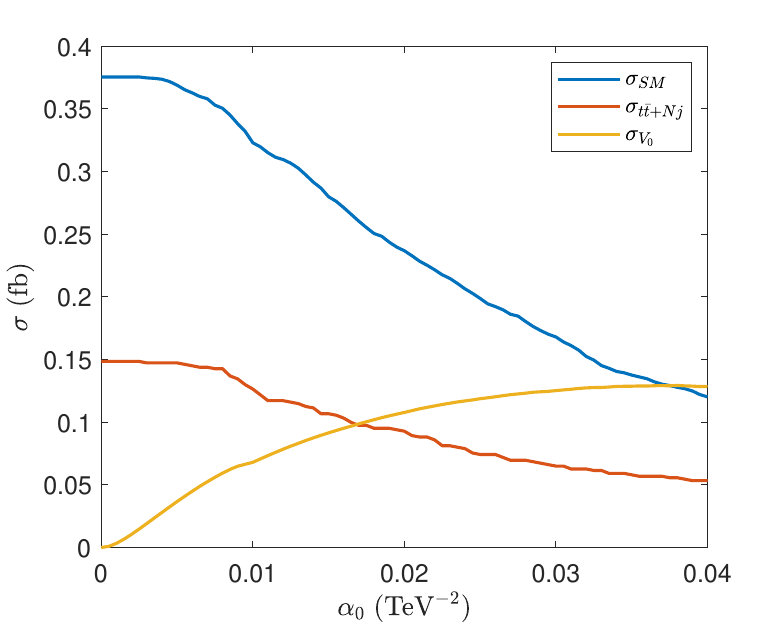}
\includegraphics[width=0.48\textwidth]{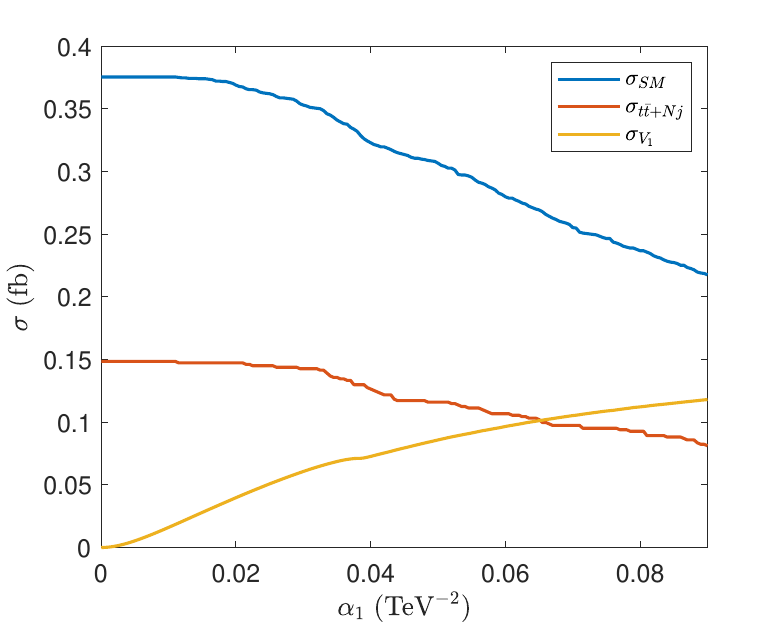}\\
\includegraphics[width=0.48\textwidth]{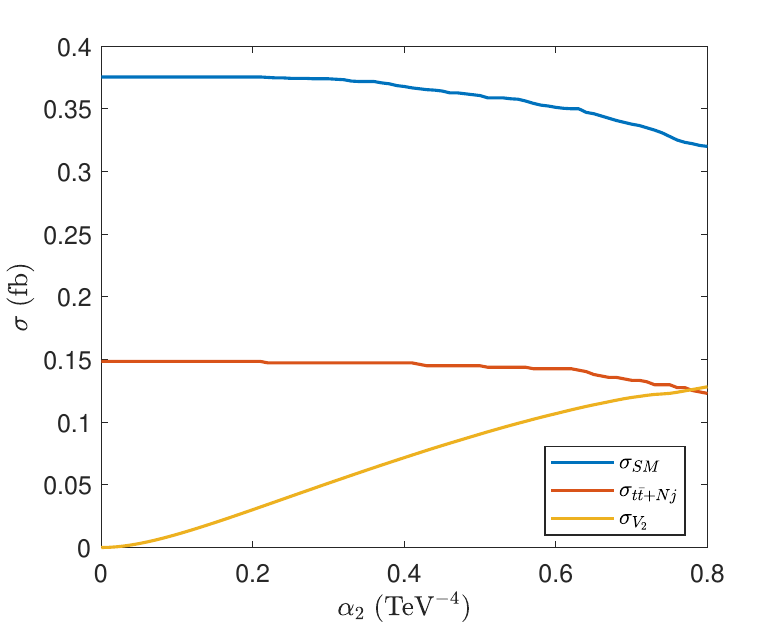}
\includegraphics[width=0.48\textwidth]{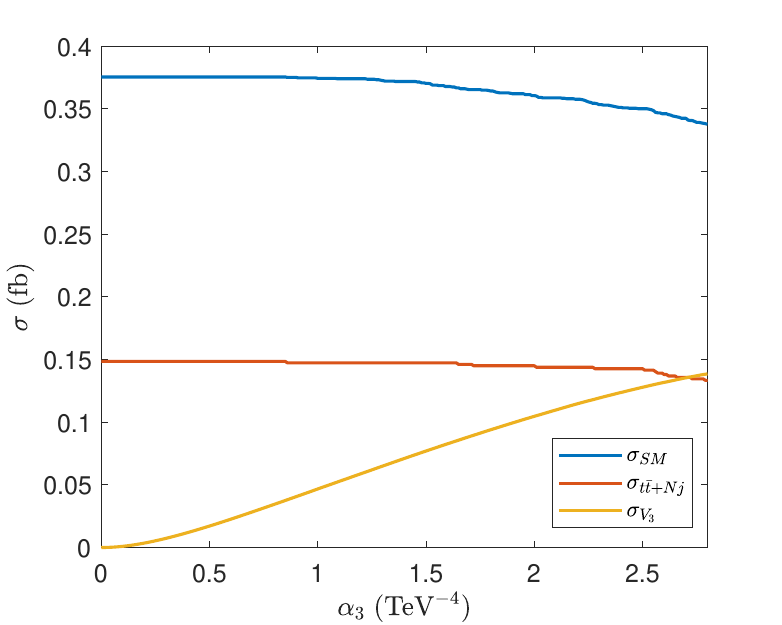}
\caption{\label{Fig:finalcs}
The cross-sections as functions of the coefficients of aQGCs after the cuts listed in Table~\ref{tab.cuts} and after the unitarity bounds.}}
\end{figure}

We scan the parameter spaces larger than the constraints listed in Table~\ref{tab.1} because the unitarity bounds have significant suppressive effects.
The unitarity bounds are applied for each coefficient individually, and are applied according to Eq.~(\ref{eq.6.4}).
The cross-sections for aQGCs, the SM and $t\bar{t}$ backgrounds are denoted as $\sigma _{V_i}$, $\sigma _{SM}$ and $\sigma _{t\bar{t}+Nj}$, respectively.
After the cuts listed in Table~\ref{tab.cuts} and after the unitarity bounds, the cross-sections as functions of the coefficients are shown in Fig.~\ref{Fig:finalcs}.
Note that, although the cross-sections of the backgrounds are not functions of the coefficients of aQGCs, we compare the cross-sections of the backgrounds under different energy scales which are related with the coefficients of aQGCs, consequently, the cross-sections of the backgrounds appear to become functions of the coefficients of aQGCs.
Without the unitarity bounds, the cross-sections of the signals should be quadratic functions of the coefficients.
As we can see from Fig.~\ref{Fig:finalcs}, this is greatly changed by the unitarity bounds.

\begin{figure}[!htbp]
\centering{
\includegraphics[width=0.48\textwidth]{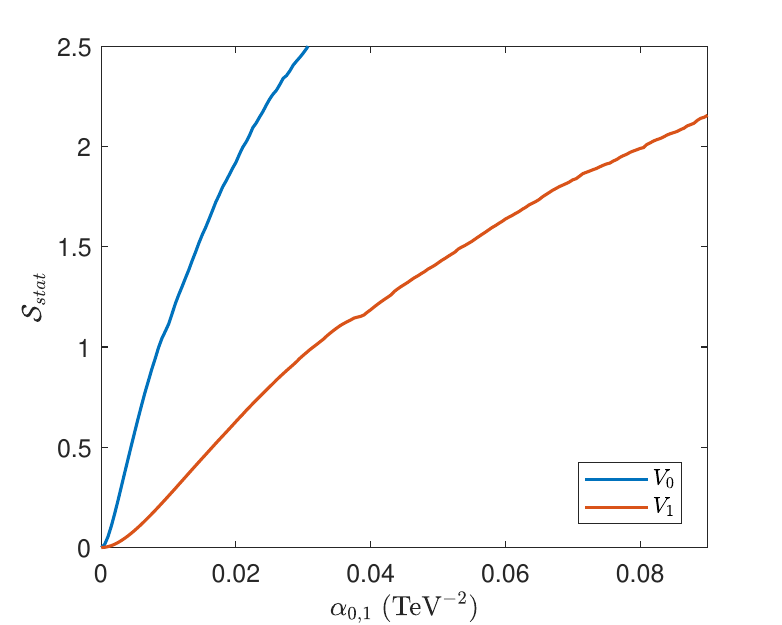}
\includegraphics[width=0.48\textwidth]{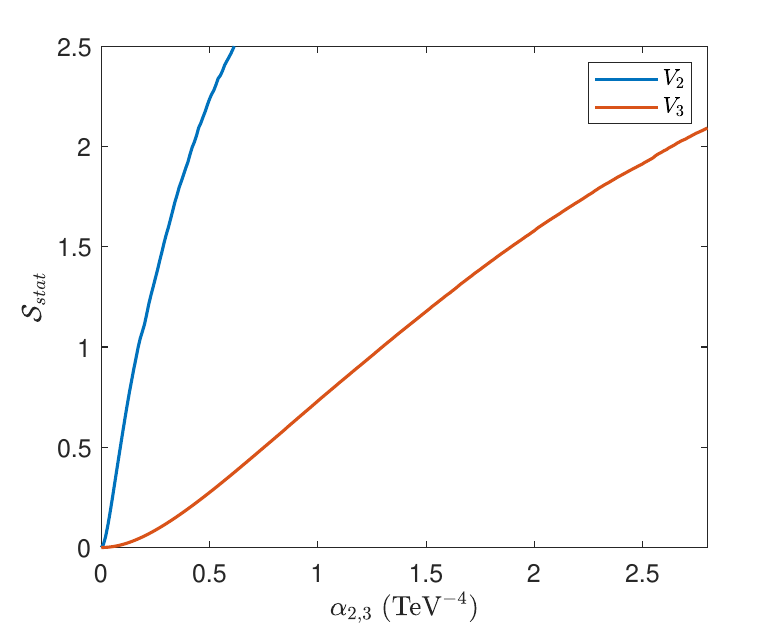}
\caption{\label{Fig:finalss}
The signal significances at $\mathcal{L}=139\;{\rm fb}^{-1}$ as functions of the coefficients of aQGCs.}}
\end{figure}

\begin{table}
\begin{center}
\begin{tabular}{c|c|c}
\hline
 & $139\;{\rm fb}^{-1}$ & $300\;{\rm fb}^{-1}$ \\
\hline
$|\alpha _0|\;({\rm TeV}^{-2})$ & $<0.021$ & $<0.013$ \\
$|\alpha _1|\;({\rm TeV}^{-2})$ & $<0.081$ & $<0.048$ \\
$|\alpha _2|\;({\rm TeV}^{-4})$ & $<0.42$ & $<0.26$\\
$|\alpha _3|\;({\rm TeV}^{-4})$ & $<2.63$ & $<1.72$\\
\hline
\end{tabular}
\end{center}
\caption{\label{Tab:constraints}The expected constraints on the coefficients of the anomalous $\gamma\gamma WW$ couplings at $13\;{\rm TeV}$ with $\mathcal{L}=139\; {\rm fb^{-1}}$ and $\mathcal{L}=300\; {\rm fb^{-1}}$ when $\mathcal{S}_{stat}< 2$.}
\end{table}

$\mathcal{S}_{stat}$ for aQGCs are calculated and shown in Fig.~\ref{Fig:finalss}.
$\mathcal{S}_{stat}$ in Fig.~\ref{Fig:finalss} is calculated at luminosity $\mathcal{L}=139\;{\rm fb}^{-1}$ which is the total luminosity at $13\;{\rm TeV}$ LHC~\cite{atlas139,139fb}.
It can be found in Fig.~\ref{Fig:finalss} that the process $pp\to jj \ell^+\ell^-\nu\bar{\nu}$ is sensitive to the $V_{0,2}$ vertices.
The expected constraints are calculated assuming the signals of aQGCs are not observed with $\mathcal{S}_{stat}\geq 2$, which are shown in Table~\ref{Tab:constraints}.
The results for possible future LHC luminosity $\mathcal{L}=300\;{\rm fb}^{-1}$~\cite{HLLHC} are also shown in Table~\ref{Tab:constraints}.
A comparison of Tables~\ref{tab.1} and \ref{Tab:constraints} shows that, except for $V_2$, the constraints at $\mathcal{L}=139\;{\rm fb}^{-1}$ in Table~\ref{Tab:constraints} is a bit less stringent, while the constraints in Table~\ref{tab.1} is given at $\mathcal{L}=35.9\;{\rm fb}^{-1}$ by studying the production of $W\gamma$.
The main reason is that results in Table~\ref{tab.1} do not take into account unitarity bounds.
The fact that constraints with unitarity bounds considered are significantly less stringent were also observed in the studies using `clipping' method~\cite{eventclipping1}.
From the results in Figs.~\ref{Fig:finalcs}, \ref{Fig:finalss} and Table~\ref{Tab:constraints}, one can see that when the unitarity bounds are applied, it is very important to increase the luminosity in order to narrow down the coefficient spaces.
This is the problem of the narrow `EFT triangles' which has been pointed out in previous studies~\cite{mo1,efttraingle2,efttraingle3},
and it has been suggested that multi-operator analysis and combination of different processes are also important.
We shall emphasis that, the above arguments are based on the pessimistic assumption that NP signals will not be discovered.
This in turn just shows the importance of the high-energy region, where plenty room has been left for the discovery of new resonances.

\section{\label{level7}Summary}

In the study of the SMEFT, the energy scale of a process is an important parameter.
However, reconstruction of the energy scales for processes at the LHC is difficult when there are two neutrinos in final states.
The energy scale of the sub-process $\gamma\gamma \to W^+W^-$ in the VBS process $pp\to jj \ell^+\ell ^- \nu\bar{\nu}$ is such a case.
In this paper, we study the contribution of aQGCs in the process $pp\to jj \ell^+\ell ^- \nu\bar{\nu}$ with the focus on the energy scale of the sub-process $\gamma\gamma \to W^+W^-$.
The method we are using is the ANN.

We show that ANN, as a technique that has proven itself in several areas of HEP, is powerful when studying $\hat{s}$ of the process $pp\to jj \ell^+\ell ^- \nu\bar{\nu}$.
The results of the ANNs are much better than the approximation derived from kinematic analysis.
With the help of ANNs, we investigate the information about $\hat{s}$ hidden in the final state.
It can be shown that, the importance of different sectors can be ordered as $p^{\ell^{\pm}} > {\bf p}_T^{\rm miss}> p^{\rm jet}$.
Apart from that, which coupling is being studied, and the collision energy are two pieces of information that are hardly used.

With the help of the ANN approach, we find another approximate formula for $\hat{s}$ which is a function of three variables $\theta _{\ell\ell}$, $E_{\ell^+}$ and $E_{\ell ^-}$ and contains only five fitting parameters, as presented in Eq.~(\ref{eq.5.3}).
Eq.~(\ref{eq.5.3}) has comparable accuracy as the ANN trained with 4-momenta of charged leptons which has $9075$ fitting parameters, and is more understandable than the ANN.
In addition, Eq.~(\ref{eq.5.3}) is much better than the approximation derived from kinematic analysis.

The unitarity bounds and the signal significances of aQGCs are also studied in this paper.
It can be shown that, $\hat{s}$ reconstructed by the ANN approach serves as an observable powerful in discriminating the signal events from the backgrounds.
With $\hat{s}$, the unitarity bounds can be applied.
The unitarity bounds have significant suppressive effects, and therefore are necessary.
With unitarity bounds applied, the cross-sections and the signal significances of aQGCs are studied.
The expected constraints at $\mathcal{L}=139$ and $300\;{\rm fb}^{-1}$ are obtained.
The constraints from the process $pp\to jj\ell^+\ell^-\nu\bar{\nu}$ can contribute to the combined limits.

\section*{ACKNOWLEDGMENT}

\noindent
This work is supported in part by the National Natural Science Foundation of China under Grants No.11905093 and No.12047570, the Natural Science Foundation of the Liaoning Scientific Committee (No.2019-BS-154) and the Outstanding Research Cultivation Program of Liaoning Normal University (No.21GDL004).

\bibliography{unitaryAAWW}
\bibliographystyle{JHEP}

\end{document}